\documentclass[aps, prb, reprint, twocolumn, longbibliography,floatfix]{revtex4-2}
\usepackage[T1]{fontenc}           
\usepackage[british]{babel} 
\usepackage{amsmath}
\usepackage{mathtools}
\usepackage{xcolor}
\usepackage{graphicx} 
\usepackage{comment}
\usepackage{hyperref}

\renewcommand{\vec}[1]{\mathbf{#1}}

\DeclareMathOperator{\sgn}{sgn}

\renewcommand{\Re}{\mathop{\mathrm{Re}}\nolimits}
\renewcommand{\Im}{\mathop{\mathrm{Im}}\nolimits}
 
\newcommand{\grad}{\boldsymbol{\nabla}}

\begin{document}
\title{Scattering approach to near-field radiative heat transfer}
\author{Matthias H\"ubler}
\affiliation{Fachbereich Physik, Universit\"at Konstanz, 78457 Konstanz, Germany}

\author{Denis M. Basko}
\affiliation{Univ. Grenoble Alpes, CNRS, LPMMC, 38000 Grenoble, France}

\author{Wolfgang Belzig}
\affiliation{Fachbereich Physik, Universit\"at Konstanz, 78457 Konstanz, Germany}

\date{\today}

\begin{abstract}
    We formulate the problem of near-field radiative heat transfer as an effective quantum scattering theory for excitations of the matter. Built from the same ingredients as the semiclassical fluctuational electrodynamics, the standard tool to handle this problem, our construction makes manifest its relation to the Landauer-B\"uttiker scattering framework, which appears only implicitly in the fluctuational electrodynamics.
    We show how to construct the scattering matrix for the matter excitations and give a general expression for the energy current in terms of this scattering matrix.
    We show that the energy current has an important non-dissipative contribution that can dominate the finite-frequency noise while being absent in the average current. Our construction provides a unified description of near-field radiative heat transfer in diverse physical systems.
\end{abstract}

\maketitle

\section{Introduction}
Mesoscopic transport of electrons, phonons, and photons is associated with quantum-mechanical scattering processes \cite{Landauer:1987,Angelescu:1998,Pekola:2021}. The Landauer-B\"uttiker scattering approach~\cite{Imry:2001, Nazarov:2009}
expresses average currents between two or more particle reservoirs in terms of the particle scattering matrix.
Moreover, the scattering approach gives a general prediction for current noise~\cite{Buettiker:1992}, which is known to contain additional valuable information beyond the average signal~\cite{Landauer:1998}. This is widely utilized in the field of electronic quantum transport, where current noise is employed to characterize conduction mechanisms and the nature of charge carriers~\cite{Blanter:2000}. Besides the field of electronic transport, a Landauer-type formula for the average current is also found in quantum heat transport conveyed by phonons or photons~\cite{Angelescu:1998, Ojanen:2007}.

Near-field radiative heat transfer (NFRHT) occurs between closely spaced bodies and is mediated by electromagnetic field rather than carried by particles. The natural and common tool for its description is fluctuational electrodynamics (FED)~\cite{Rytov:1953, Rytov:1989}, which also results in a Landauer-type formula for the average heat current~\cite{Polder:1971, Joulain:2005, Volokitin:2007, Bimonte:2017}. The analogy between the FED formulation of NFRHT and Landauer-B\"uttiker approach was pushed further in Refs.~\cite{Biehs:2010, Yap:2017}, where thermal reservoirs and transmission channels were identified. Still, we are not aware of any work where scattering states for reservoir excitations and the corresponding scattering matrix would be constructed for the NFRHT problem.
The need for such construction becomes especially pressing in the context of recent studies of heat-current noise in NFRHT~\cite{Golubev:2015, Biehs:2018, Tang:2018, Herz:2020, Wise:2022}. It has been noted~\cite{Wise:2022} that the NFRHT noise spectrum does not fit the general form obtained from the scattering approach for bosonic particles in Ref.~\cite{Buettiker:1992}. It is therefore important to clarify what type of scattering problem NFRHT corresponds to.

\begin{figure}
    \centering
     \includegraphics[width=0.45\textwidth]{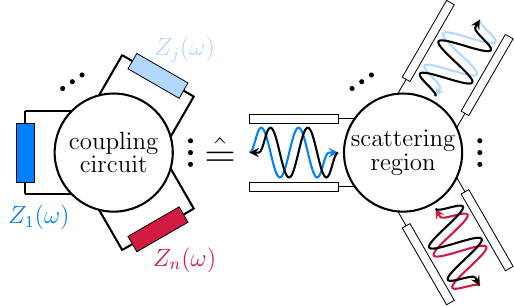}
     \caption{Left: Circuit representation of near-field radiative heat transfer (NFRHT) between thermal reservoirs at different temperatures. The reservoirs are modeled as resistors, characterized by their frequency-dependent impedances, $Z_j(\omega)$. Right: An equivalent scattering representation of NFRHT, where transmission lines (TL's) couple through a scattering region. Incoming TL excitations are either reflected back or transmitted to other TL's through the scattering region.}
     \label{fig:network}
\end{figure}

In this work, we show how to formulate the quantum-mechanical scattering problem for NFRHT, focusing on its circuit version~\cite{Pascal:2011} for simplicity.
Effective circuits of lumped elements (resistors, capacitors, inductors) can be used to model NFRHT between nanoscale objects at low temperatures~\cite{Meschke:2006, Ojanen:2008, Pascal:2011, Wise:2020role}, when the full set of Maxwell's equations separates into electrostatic and magnetostatic sectors with few relevant modes. In the circuit picture, thermal reservoirs associated with different bodies at different temperatures are represented as dissipative resistors, which include fluctuating current sources (Nyquist-Johnson noise at the corresponding temperature) and are coupled by non-dissipative linear circuit elements (capacitors and inductors, modeling electrostatic and magnetostatic coupling, respectively), as schematically represented in Fig.~\ref{fig:network}. Instead of Maxwell's equations, one solves Kirchhoff's laws, and all geometrical details of the structure are encoded in a few circuit parameters~\cite{Wise:2020role}. Such circuit models are more tractable than the standard Maxwellian FED, while being conceptually equivalent.

Using an equivalent representation of each reservoir by a microwave transmission line (TL) hosting bosonic excitations, we construct scattering states for these excitations. The corresponding scattering matrix can be determined from the linear circuit equations (Kirchhoff's laws), just like the response to fluctuating sources in FED. The scattering theory enables one to determine arbitrary high-order correlation functions of the NFRHT energy current. We obtain a general expression for the energy current, which extends B\"uttiker's formula for the particle current~\cite{Buettiker:1992, Blanter:2000}. We find that frequency dependence of reservoir's impedance results in an additional reactive contribution to the energy current, which has no counterpart in B\"uttiker's theory for particle current. This reactive contribution is absent in the average energy current, but can give a dominant contribution to its finite-frequency noise.

The article is organized as follows. First, in Sec.~\ref{sec:RestoTL}, we consider heat transport in electric circuits and construct the scattering states for each thermal reservoir. 
Sec.~\ref{sec:S} introduces  the scattering matrix and establishes the connection between the scattering approach and circuit fluctuational electrodynamics. 
Next, in Sec.~\ref{sec:ECurr}, we construct the energy current operator and examine the average dissipated power. Then, in Sec.~\ref{sec:HCurrNoise}, we calculate the heat-current noise spectrum and contrast it with B\"uttiker’s formula, illustrating the differences using two dielectrics coupled by surface phonon–polaritons. 
Sec.~\ref{sec:ExtendedBodies} extends the circuit-based scattering approach to radiative heat transfer between spatially extended bodies.
Thereafter, in Sec.~\ref{sec:Conclusion} we present our conclusions.
Several appendices contain technical details. In Appendix~\ref{App:ResistorClassicalDrive}, we discuss the response of a resistor to a classical drive. Appendix~\ref{App:SimpleCirc} contains the calculation of heat-current fluctuations in simple circuits. In Appendix~\ref{sec:circuit_polaritons} we present an analytical calculation of heat-current fluctuations in the effective circuit that mimics surface phonon–polaritons. Finally, in Appendix~\ref{App:GreenSurfacePolariton}, we present a microscopic non-equilibrium Green’s-function calculation of heat-current fluctuations mediated by surface phonon–polaritons for two dielectrics in the planar geometry, and emphasize its relation to the effective-circuit description.

\begin{figure}
    \centering
     \includegraphics[width=\columnwidth]{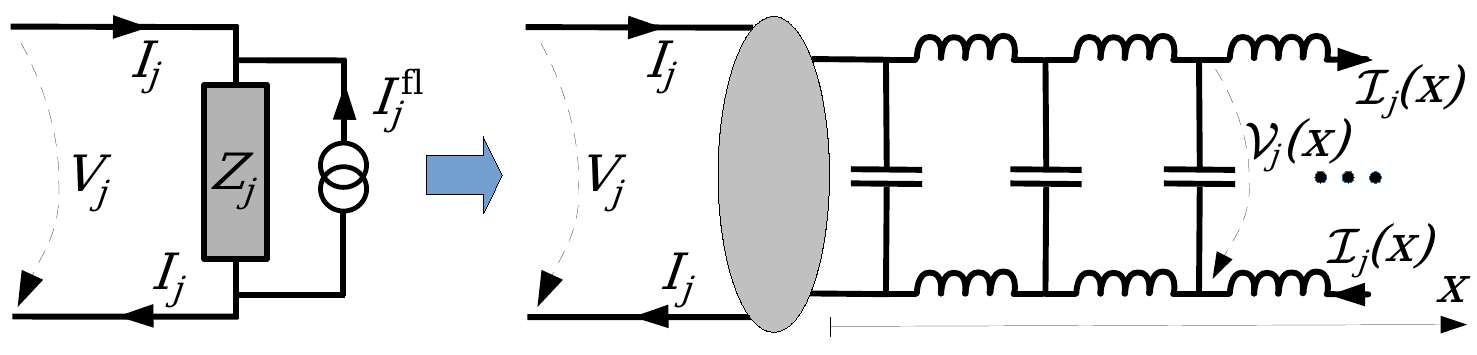}
     \caption{Representation of a resistor $Z_j(\omega)$ by an extended semi-infinite TL, modeled by a sequence of infinitesimal inductors and capacitors, connected to the rest of the circuit via a linear 4-terminal element (``filter'') which transforms a constant impedance of the TL into the given impedance $Z_j(\omega)$.}
     \label{fig:resistor}
\end{figure}

\section{Thermal reservoir as a transmission line}
\label{sec:RestoTL}
We employ the mesoscopic approach of circuit quantum electrodynamics \cite{Yurke:1984,Vool:2017}, directly quantizing circuit fluxes and charges, which yield currents and voltages.
It is equivalent to the Green’s function approach in the random-phase approximation~\cite{Tang:2018,Wise:2022}, where currents and voltages emerge as collective variables of electronic or ionic motion. 
Our main step is to represent each resistor~$j$ having an arbitrary complex impedance $Z_j(\omega)$ as a semi-infinite TL with some real constant impedance~${z}_j$ (thus purely dissipative), combined with a purely reactive element which yields the required total impedance $Z_j(\omega)$~\cite{Pauli:1989}. The TL is characterized by the inductance $\ell_j$ and capacitance $\varsigma_j$ per unit length, and can be viewed as a chain of infinitesimal unit cells of length $\Delta{x}\to0$, each unit cell containing two inductances $\ell_j\,\Delta{x}/2$ and a capacitance $\varsigma_j\,\Delta{x}$ (Fig.~\ref{fig:resistor}). The TL's impedance is ${z}_j=\sqrt{\ell_j/\varsigma_j}$ and the wave velocity $c_j=1/\sqrt{\ell_j\varsigma_j}$. The voltage and current profiles
\begin{subequations}\label{eqs:IVxt}\begin{align}
&\hat{\mathcal{V}}_j(x,t)=\int\limits_0^\infty{d}\omega
\sqrt{\frac{\hbar\omega{z}_j}{4\pi}}\sum_{\sigma=\pm}\left[\hat{a}_{j,\omega}^\sigma{e}^{-i\omega(t-\sigma{x}/c_j)}+\mbox{h.c.}\right],\\
&\hat{\mathcal{I}}_j(x,t)=\int\limits_0^\infty{d}\omega
\sqrt{\frac{\hbar\omega}{4\pi{z}_j}}\sum_{\sigma=\pm}\left[\sigma\hat{a}_{j,\omega}^\sigma{e}^{-i\omega(t-\sigma{x}/c_j)}+\mbox{h.c.}\right],  
\end{align}\end{subequations}
are expressed in terms of bosonic creation/annihilation operators $\hat{a}^{\sigma\,\dagger}_{j,\omega},\hat{a}^\sigma_{j,\omega}$, for left- ($\sigma=-$) and right- ($\sigma=+$) propagating excitations at frequency~$\omega$ with the commutator $[\hat{a}^{\sigma}_{j,\omega},\hat{a}^{\sigma\, \dagger}_{k,\omega'}]=\delta_{jk}\delta(\omega-\omega')$. The scattered excitations, $\sigma=+$, are determined by the incident excitations and by the coupling circuit, so the operators $\hat{a}^+_{j,\omega}$ are related to $\hat{a}^-_{k,\omega'}$. Quantization of fluxes and charges in the circuit constitutes the circuit analog of quantizing the electric and magnetic fields in the Maxwellian electrodynamics.

At $x=0$ the TL is connected to a fictitious linear reciprocal 4-terminal element (``filter'') in order to reproduce the required impedance $Z_j(\omega)$. The filter transforms voltage and current Fourier components as~\cite{Pauli:1989} 
\begin{align}
\begin{pmatrix}
\hat{V}_j(\omega) \\ \hat{I}_j(\omega)
\end{pmatrix}
= 
\begin{pmatrix}
\Re\frac{Z_j(\omega)}{\lambda_j(\omega)} & i\Im\frac{Z_j(\omega)}{\lambda_j(\omega)} \\
i\Im\frac1{\lambda_j(\omega)} & \Re\frac1{\lambda_j(\omega)}
\end{pmatrix}
\begin{pmatrix}
z_j^{-1/2}\hat{\cal V}_j(0,\omega) \\ z_j^{1/2}\hat{\cal I}_j(0,\omega)
\end{pmatrix}.
\label{eq:filtertransform}
\end{align}
The function $\lambda_j(\omega)$ is determined by the physical realization of the filter (the so-called network synthesis~\cite{Balabanian:1969,Pauli:1989}). It must satisfy $|\lambda_j(\omega)|^2=\Re Z_j(\omega)$. To respect causality, $\lambda_j(\omega)$ and $1/\lambda^*_j(\omega)$ must be holomorphic for $\Im\omega > 0$. These conditions do not uniquely determine $\lambda_j(\omega)=\sqrt{\Re{Z}_j(\omega)}\,e^{i\vartheta_j(\omega)}$, leaving an arbitrariness in the phase $\vartheta_j(\omega)$. For example, one may include a piece of transmission line of length~$l$ into the filter, which amounts to $\vartheta_j\to\vartheta_j+\omega{l}/c_j$. We do not need the explicit construction, since measurable quantities are phase-independent. 

The resulting expressions of the voltage and current at the resistor~$j$ in terms of the bosonic mode operators,
\begin{subequations}\label{eqs:IV=modes}\begin{align}
\hat{V}_j(t)={}&{}\int_0^\infty{d}\omega
\sqrt{\frac{\hbar\omega}{4\pi}}\left[ \frac{Z_j(\omega)}{\lambda_j(\omega)}\,\hat{a}_{j,\omega}^++\frac{Z_j^*(\omega)}{\lambda_j^*(\omega)}\,\hat{a}_{j,\omega}^- \right]{e}^{-i\omega{t}} \nonumber\\
&{}+\mbox{h.c.},
\label{eq:Vt=modes}\\
\hat{I}_j(t)={}&{}\int_0^\infty{d}\omega
\sqrt{\frac{\hbar\omega}{4\pi}}
\left[\frac{1}{\lambda_j(\omega)}\,\hat{a}_{j,\omega}^{+}-\frac{1}{\lambda^*_j(\omega)}\,\hat{a}_{j,\omega}^{-}\right]{e}^{-i\omega{t}} \nonumber\\
&{}+\mbox{h.c.},
\label{eq:It=modes}
\end{align}\label{eq:IV=modes}\end{subequations}
will be our main tool in the following.
We investigate the heat exchange between the resistors while maintaining their temperatures at $T_j$. Accordingly, the reservoirs are assumed to have thermal populations $\langle \hat{a}^{-\,\dagger}_{j,\omega}\,\hat{a}^-_{k,\omega'}\rangle=\delta_{jk} n_j(\omega) \delta(\omega-\omega')$, where $n_j(\omega)=1/(e^{\hbar\omega/T_j}-1)$ being the Bose-Einstein distribution
(the Boltzmann constant $k_\text{B}=1$). 
In thermal equilibrium, the expectation values $\langle \hat{a}^{-}_{j,\omega} \rangle$ vanish, resulting in zero average currents and voltages. However, other quantum states of the TL's can give rise to nonzero average currents and voltages.
Applying a classical voltage drive to the resistor means imposing $\hat{V}_j(t)=V_\text{ext}e^{-i\omega_\text{ext}t}+\mbox{c.c.}$. Then we obtain $\hat{a}_{j,\omega}^+ = V_\text{ext}[\lambda_j(\omega)/Z_j(\omega)]\sqrt{4\pi/(\hbar\omega)}\delta(\omega-\omega_\text{ext}) - [Z_j^*(\omega)/Z_j(\omega)]e^{2i\vartheta_j(\omega)}\hat{a}_{j,\omega}^-$ from Eq.~(\ref{eq:Vt=modes}), and
Eq.~(\ref{eq:It=modes}) yields $\langle\hat{I}_j(t)\rangle = V_\text{ext}e^{-i\omega_\text{ext}t}/Z_j(\omega_\text{ext})+\mbox{c.c.}$, consistent with the definition of~$Z_j(\omega)$ (see Appendix~\ref{App:ResistorClassicalDrive}).

\section{Scattering matrix}
\label{sec:S}
The operators $\hat{a}_{j,\omega}^+$ of scattered excitations are related to those of incident excitations. In a linear coupling circuit without external drives, this relation must be linear and frequency-conserving, $\hat{a}^+_{j,\omega} = \sum_k S_{jk}(\omega) \hat{a}^-_{k,\omega}$, where $S_{jk}(\omega)$ is the scattering matrix~\footnote{
{T}ypically, the coupling circuit does not conserve the number of excitations; indeed, electrostatic/magnetostatic coupling enters the circuit Hamiltonian via charges/fluxes which involve linear combinations of creation and annihilation operators. However, this non-conservation may occur only transiently in the scattering region, but not in the asymptotic scattering states. Mixing creation and annihilation operators in the scattering relation would allow for production of excitation pairs out of vacuum, which is forbidden in a stable harmonic system without external drives.
}.
It must be holomorphic for $ \Im (\omega) > 0$ by causality; otherwise, the scattered excitation $\hat{a}^+_j(t) = \sum_k \int dt' S_{jk}(t-t') \hat{a}^-_k(t') $ would depend on $\hat{a}^-_j(t')$ in the future. Since the coupling circuit is lossless (all dissipative elements must be included as reservoirs), the scattering matrix must be unitary, $\sum_k S_{kj}^*(\omega)S_{kl}(\omega)=\delta_{jl}$~\cite{Pozar:2021}. 
It is sufficient to solve the linear circuit equations (Kirchhoff's laws), with no need to specify the circuit Hamiltonian, to find the scattering matrix
\begin{align}
 S_{jk}(\omega)=& \frac{a^+_{j,\omega}}{a^-_{k,\omega}} \bigg |_{\substack{a^-_{l,\omega}=0,\\~~~l \neq k}}=\frac{\lambda_j(\omega)}{\lambda_k^*(\omega)}\frac{I^+_{j}(\omega)}{I^-_{k}(\omega)} \bigg |_{\substack{I^-_{l}(\omega)=0, \\ ~~~~~l \neq k}} \label{eq:S}
\end{align}%
where $I^-_j(\omega)\equiv[V_j(\omega)-Z_j(\omega)I_j(\omega)]/[2\Re{Z}_j(\omega)]$ is the incident current and $I^+_j(\omega)\equiv[V_j(\omega)+Z^*_j(\omega)I_j(\omega)]/[2\Re{Z}_j(\omega)]$ the scattered current.

In the scattering approach, fluctuations emerge from the incident TL excitations, whereas in circuit FED \cite{Pascal:2011}, they originate from fictitious fluctuating current sources $I^\text{fl}_j(t)$ attached in parallel to each resistor (Fig. \ref{fig:resistor}). $I^\text{fl}_j(t)$ represents the
Johnson-Nyquist noise of the resistor, dictated by the fluctuation-dissipation theorem, and induces the fluctuating voltage $V_j(\omega)=Z_j(\omega) [I_j(\omega)+I^{\text{fl}}_j(\omega)]$. In the scattering approach, Eqs.~\eqref{eqs:IV=modes} lead to the fluctuating voltage $V_j(\omega)=Z_j(\omega)I_j(\omega)+[2 \Re Z_j(\omega)]I^-_j(\omega)$. Thus, the incident current plays the same role as the fluctuating current, and both approaches induce the identical voltage if $Z_j(\omega)\,I^\text{fl}_j(\omega)=[2\Re{Z}_j(\omega)]I_j^-(\omega)$. We recover the standard Johnson-Nyquist expression for the average anticommutator $\langle\{\hat{I}^-_j(t),\hat{I}^-_j(0)\}\rangle$. Additionally, we can apply the circuit FED approach~\cite{Pascal:2011} to determine the scattering matrix.
The fluctuating current source $I^{\text{fl}}_j(\omega)$ at the resistor~$k$ induces a voltage $V_j(\omega)=\sum_k \zeta_{jk}(\omega)\,I_k^\text{fl}(\omega)$ on resistor $j$, where the matrix $\zeta_{jk}(\omega)$ is found by using Kirchhoff's laws. (The matrix $\zeta_{jk}(\omega)$ should not be confused with the impedance matrix which relates the total currents, not the fluctuating sources.) 
Then, using Eqs.~\eqref{eqs:IV=modes}, \eqref{eq:S} and the relation $Z_j(\omega)\,I^\text{fl}_j(\omega)=[2\Re{Z}_j(\omega)]I_j^-(\omega)$, we can express the scattering matrix as 
\begin{equation}
 {S}_{jk}(\omega)=\frac{2\lambda_j(\omega)\lambda_k(\omega)}{Z_j(\omega)\,Z_k(\omega)}\,\zeta_{jk}(\omega) - \delta_{jk}\,\frac{\lambda_j(\omega)Z_j^*(\omega)}{\lambda^*_j(\omega)Z_j(\omega)}.  
\end{equation}
In a reciprocal circuit, the matrix $\zeta_{jk}(\omega)$ is symmetric, and so is ${S}_{jk}(\omega)$. Non-reciprocity can be included via non-reciprocal circuit elements such as circulators~\cite{Pozar:2021}.

\section{Energy current}
\label{sec:ECurr}
We define the power, injected into the $j$-th resistor from the rest of the circuit, as $\hat{\cal P}_j(t)\equiv \{\hat{I}_j(t),\hat{V}_j(t)\}/2$. For its time average, Eqs.~(\ref{eqs:IV=modes}) give the natural expression
\begin{equation}
    \int_{-\tau/2}^{\tau/2}\frac{dt}\tau\,\hat{\cal P}_j(t)
    \mathop{\sim}_{\tau\to\infty}\frac{1}\tau\int_0^\infty{d}\omega\,\hbar\omega\sum_{\sigma=\pm}\sigma\, \hat{a}_{j,\omega}^{\sigma\,\dagger}\hat{a}_{j,\omega}^{\sigma}, 
\end{equation}
which is determined by the number operator of incident and scattered excitations, each carrying energy $\hbar \omega$. The same expression is obtained for the time average of the power $\{\hat{\cal I}_j(0,t),\hat{\cal V}_j(0,t)\}/2$ injected into the $j$-th TL from Eqs~(\ref{eqs:IVxt}). Thus, the introduced filter is indeed non-dissipative, although it can accumulate some energy for a finite interval of time, as any reactive element.

We express the dissipated power (energy current) operator in terms of the incident creation/annihilation operators $\hat\alpha_{j,\omega>0}\equiv\hat{a}_{j,\omega}^-$ and $\hat\alpha_{j,\omega<0}\equiv\hat{a}_{j,\omega}^{-\dagger}$, which then reads
\begin{subequations}\label{eqs:power}
\begin{align}
    \hat{\cal P}_j(t) =
    {}&{} -\iint_{-\infty}^\infty \frac{d \omega\, d \omega'}{4 \pi}\, \hbar\sqrt{|\omega\omega'|} \,e^{i (\omega-\omega')t} \nonumber \\ 
    {}&{} \times \sum_{k,l} A_{kl}^j(\omega,\omega') \, \hat\alpha_{k,\omega}^\dagger\hat\alpha_{l,\omega'},
    \label{eq:ECoperator}
\end{align}
where we defined $S_{ij}(\omega<0)=S_{ij}^*(\omega)$, and the matrix $A_{kl}^j(\omega,\omega')$ is given by
\begin{align}
A^j_{kl}(\omega,\omega')= {}&{} - \frac{Z_j^*(\omega)+Z_j(\omega')}{2 \lambda_j^*(\omega) \lambda_j(\omega')}\,S_{jk}^*(\omega)\,S_{jl}(\omega')
\nonumber \\  {}&{} - 
\frac{Z_j(\omega)-Z_j(\omega')}{2\lambda_j(\omega)\lambda_j(\omega')}\,\delta_{jk}S_{jl}(\omega')
\nonumber \\  {}&{} + 
\frac{Z_j^*(\omega)-Z_j^*(\omega')}{2\lambda_j^*(\omega)\lambda_j^*(\omega')}\,S_{jk}^*(\omega)\,\delta_{jl}
\nonumber \\  {}&{} +\frac{Z_j(\omega)+Z_j^*(\omega')}{2 \lambda_j(\omega)\lambda_j^*(\omega')}\,\delta_{jk}\delta_{jl}.
\label{eq:Amatrix}
\end{align}
\end{subequations}
These expressions can be viewed as the extension of the general B\"uttiker's formula~\cite{Buettiker:1992, Blanter:2000} to the case of energy current, and represent the main formal result of our work. The energy current operator does not conserve the particle number, as it contains terms $\hat a_{j,\omega}^{-\dagger}\hat a^{-\dagger}_{k,\omega'}, \hat a_{j,\omega}^{-}\hat a^{-}_{k,\omega'}$. This highlights a difference between mesoscopic electronic transport and radiative energy transport, where energy is carried by the electromagnetic field rather than by particles \cite{Khondker:1988,VanHaeringen:1990}, leading to these additional terms. 

To understand the origin of the difference between Eqs.~(\ref{eqs:power}) and the simpler B\"uttiker's formula, we note that the latter is recovered if instead of the power flowing into the whole impedance $Z_j(\omega)$, one considers the power flowing into its TL part; it is obtained by setting $Z_j,Z_j^*\to z_j$ and $\lambda_j \to \sqrt{z_j}$ in Eq.~\eqref{eq:Amatrix} with the same scattering matrix. This B\"uttiker-like energy current $\hat{\cal P}^B_j(t)$ matches that in Refs.~\cite{Buettiker:1992,Blanter:2000}, where the effective particle velocity $v_j\propto1/z_j$. Here, the frequency dependence of $Z_j(\omega)$ necessarily implies $\Im{Z}_j(\omega)\neq0$ due to Kramers-Kronig relations and results in a reactive component of the reservoir, where energy can be stored temporarily. The reactive contribution $\hat{\cal P}_j-\hat{\cal P}_j^B$ has no counterpart in the particle current since particle velocity must be real, even if energy-dependent. 

Equations (\ref{eqs:power}) result in a Landauer-type expression for the average dissipated power (or average energy current),
\begin{align}
\langle \hat{\cal P}_j \rangle = \sum_k  \int_0^\infty \frac{d \omega}{2 \pi}\, \hbar\omega |S_{jk}(\omega)|^2 \left[n_k(\omega)- n_j(\omega)\right],
\end{align}
well-known in the context of NFRHT; here we see how the transmission coefficient $|S_{jk}(\omega)|^2$, usually introduced phenomenologically, arises explicitly from the scattering framework.
The added value of Eqs.~(\ref{eqs:power}) is that they enable one to calculate all higher-order correlation functions of the energy current, just like the B\"uttiker's formula for the electric current. Indeed, the incident excitations being in the thermal state, arbitrary averages of the $\hat\alpha_{j,\omega}$ operators can be evaluated from the pair average $\langle\hat\alpha_{j,\omega}\hat\alpha_{k,\omega'}\rangle = \delta_{jk}\delta(\omega+\omega')\,n_j(\omega')\sgn\omega'$ using Wick's theorem~\footnote{Wick's theorem guarantees that energy current correlators do not depend on the phases $\vartheta_j(\omega)$. Indeed, $S_{jk}(\omega)\propto{e}^{i\vartheta_j(\omega)+i\vartheta_k(\omega)}$ and $A_{kl}^j(\omega,\omega')\propto{e}^{-i\vartheta_k(\omega)+i\vartheta_l(\omega')}$, so each annihilation/creation operator~$\hat\alpha_{j,\omega}$ in Eq. \eqref{eq:ECoperator} is accompanied by the corresponding phase factor. Since $\vartheta_j(\omega)=-\vartheta_j(-\omega)$, the phase factors cancel out upon averaging.}. Note that the correlations $\langle \hat a_{j,\omega}^{-\dagger}\hat a^{-\dagger}_{k,\omega'}\rangle , \langle \hat a_{j,\omega}^{-}\hat a^{-}_{k,\omega'}\rangle $ vanish, ensuring that all observables depend solely on the occupation number. Power correlations cannot be computed in the semiclassical FED framework, which contains only the average anticommutator of fluctuating currents, since correlators of order~2 and higher also involve commutators.

\section{Heat-current noise spectrum}
\label{sec:HCurrNoise}
Energy current correlations between two reservoirs $j$ and $k$ are given by
\begin{align}
W_{jk}(\Omega) = {}&{} \int_{-\infty}^\infty dt\, e^{i \Omega t} \left[\frac12\left\langle \{\hat{\cal P}_j(t), \hat{\cal P}_k(0) \} \right\rangle -\langle\hat{\cal P}_j\rangle\langle\hat{\cal P}_k\rangle\right]\nonumber \\
= {}&{}\sum_{l,m} \int_{-\infty}^\infty \frac{d\omega}{16 \pi} \hbar^2 \omega (\omega+\Omega) \nonumber\\
{}&{} \times A^j_{lm}(\omega,\omega+\Omega) A^k_{ml}(\omega+\Omega,\omega)  \nonumber\\
{}&{} \times\left[\coth\frac{\hbar\omega}{2 T_l}\coth\frac{\hbar(\omega+\Omega)}{2T_m}-1\right].
\label{eq:Wjk=}
\end{align}
A similar expression $W_{jk}^B(\Omega)$, matching the one in Refs.~\cite{Buettiker:1992,Blanter:2000}, can be constructed for the the correlations of energy currents $\hat{\cal P}^B_j(t)$ between TLs, by replacing  $Z_j,Z_j^*\to z_j$ and $\lambda_j \to \sqrt{z_j}$ in Eq.~\eqref{eq:Amatrix} for $A^j_{kl}(\omega,\omega')$~\footnote{
{A} minor difference from the particle current fluctuations in Refs.~\cite{Buettiker:1992,Blanter:2000} is the contribution from averages $\langle\hat\alpha_{k,\omega_1}\hat\alpha_{l,\omega_2}\rangle\langle\hat\alpha_{k',\omega_1'}^\dagger\hat\alpha_{l',\omega_2'}^\dagger\rangle$ coming from Wick's theorem. Due to the symmetry $A^j_{kl}(-\omega,-\omega')=A^j_{lk}(\omega',\omega)$, this only produces a factor of 2 in $W_{ij}(\Omega)$.
}.
In contrast to Eq.~(\ref{eq:Wjk=}), thus defined $W_{jk}^B(\Omega)$ depends on the phases~$\vartheta_j(\omega)$, unlike the average $\langle\hat{\cal P}_j^B(t)\rangle = \langle\hat{\cal P}_j(t)\rangle$.

The simple physical reason for such behavior is that a reactive element can accumulate some energy for a finite amount of time, and thus affect finite-frequency heat-current fluctuations, while a finite average current can only be accommodated by a dissipative element. Which reactive elements must be included into the reservoirs and which belong to the external circuit depends on their microscopic nature and on how the energy current is actually measured. Typically, one measures changes in electronic temperatures and its fluctuations~\cite{Karimi:2020}. For example, a small piece of metal with the Drude conductivity $\sigma(\omega)=\sigma_0/(1-i\omega\tau)$ ($\tau$ being the momentum relaxation time) can be effectively represented as a resistor and an inductor in series. A tunnel junction is equivalent to a resistor and a capacitor in parallel. However, in both cases the reactive and the dissipative components are effectively made of the same electrons, there is no way to physically separate them, so $\hat{\cal P}^B(t)$ is not measurable.

\begin{figure}
    \includegraphics[width=\columnwidth]{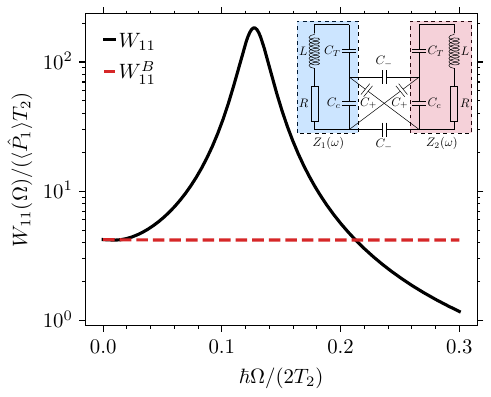}
    \caption{Heat-current noise spectrum $W_{11}(\Omega)$ (solid line) and the B\"uttiker-like contribution $W_{11}^B(\Omega)$ (dashed line), for the circuit mimicking heat transfer via surface polaritons (inset) with $L=44\:\text{pH}$, $C_T=1\:\text{aF}$, $C_c=2.3\:\text{aF}$, $C_+=1.5\:\text{aF}$, $C_-=1.0\:\text{aF}$, $R=44\:\Omega$ at $T_1\to0$, $T_2=300\:\text{K}$.}
    \label{fig:Noisepolariton_circuit}
\end{figure}

Importance of the reactive contribution to the noise is illustrated with some simple circuits in Appendix~\ref{App:SimpleCirc} and is particularly manifest in fluctuations of NFRHT between two dielectrics via coupled surface phonon-polaritons~\cite{Biehs:2018}. This physics (electric coupling between two polarization oscillators) can be mimicked by the effective circuit with two $RLC$ contours, hosting weakly damped resonances at frequency $\omega_L=\sqrt{1/(LC_T)+1/(LC_\text{c})}$, and coupled by the capacitances~$C_\pm$ (see Appendix~\ref{sec:circuit_polaritons} and Fig.~\ref{fig:Noisepolariton_circuit}). The frequency of the coupled symmetric (antisymmetric) mode is $\omega_\pm=\sqrt{1/(LC_T)+1/(LC_\text{c}+LC_\pm)}$, and their amplitude decay rate $\gamma=R/(2L)$. The Poynting vector in the vacuum gap separating the two dielectrics corresponds to the power deposited into each $RLC$ contour, which is thus treated as a dissipative bath with impedance
\begin{align}
Z(\omega) = {}&{} \left(-i\omega{C}_\text{c} - \frac{i\omega{C}_T}{1-\omega^2LC_T-i\omega{R}C_T}\right)^{-1},
\end{align}
while the capacitances $C_\pm$ make the coupling circuit.
We choose $\lambda(\omega) = [\sqrt{R}/(LC_\text{c})]/(\omega_L^2-\omega^2-2i\omega\gamma)$ and find
\begin{align}
    |S_{12}(\omega)|^2 = \frac{(2\gamma\omega)^2(\omega_+^2-\omega_-^2)^2}{\prod_{\sigma=\pm}[(\omega_\sigma^2-\omega^2)^2+(2\gamma\omega)^2]},
\end{align}
as detailed in Appendix~\ref{sec:circuit_polaritons}.
Generally, fluctuations are more important in smaller systems, for which the circuit description is also expected to be most suitable. 
In Fig.~\ref{fig:Noisepolariton_circuit} we plot $W_{11}(\Omega)$ and $W_{11}^B(\Omega)$ for the circuit parameters representative of two pieces of dielectric of volume $(30\:\text{nm})^3$, separated by a vacuum gap of size $d=30\:\mbox{nm}$: the typical capacitance $C_T=1\:\text{aF}$, $1/\sqrt{LC_T},\omega_L=(1.5,1.8)\times10^{14}\:\mbox{s}^{-1}$ (transverse and longitudinal optical phonons in SiC, respectively), $\omega_+,\omega_-=(1.7,1.6)\times10^{14}\:\mbox{s}^{-1}$, $\gamma=5\times10^{11}\:\mbox{s}^{-1}$, which fix all circuit parameters, and ${T}_2=300\:\mbox{K}\gg{T}_1$.
The reactive contribution dominates in a wide range of~$\Omega$ and has a prominent peak at $\Omega=|\omega_+-\omega_-|$. This peak corresponds to oscillations in the time domain, found in Ref.~\cite{Biehs:2018}.\\

To prove that the circuit in Fig.~\ref{fig:Noisepolariton_circuit} does indeed provide a correct model for the NFRHT via coupled surface phonon-polaritons, in Appendix~\ref{App:GreenSurfacePolariton} we present a microscopic calculation for two dielectrics separated by a planar vacuum gap, using non-equilibrium Green's functions. By comparing the two calculations, one can see that each heat conduction channel (in the planar geometry specified by the in-plane wave vector) behaves exactly as the effective circuit. This confirms the common knowledge that the circuit representation works when the electromagnetic coupling is dominated by just a few modes (which is often the case for small objects at low temperatures). At the same time, the Green's function calculation is much heavier, and the relation to the standard Landauer-B\"uttiker framework is not manifest. In a similar fashion, equivalence of scattering approach to the Green's function formalism can be established rather straightforwardly for each specific system (since the use of Green's functions is always based on a specific microscopic model with a certain Hamiltonian).

\section{Scattering problem for radiative heat transfer between extended bodies}
\label{sec:ExtendedBodies}

The scattering problem developed for heat transfer in circuits can be straightforwardly extended to the standard picture of radiative heat transport between spatially extended bodies and fields described by Maxwell's equations. Let us focus on structures described by a position-dependent local isotropic optical conductivity $\sigma(\vec{r},\omega)$ [or, equivalently, dielectric function $\varepsilon(\vec{r},\omega)=1+4\pi{i}\sigma(\vec{r},\omega)/\omega$] without spatial dispersion, which covers most of situations studied in the literature. The crucial step to construct the scattering problem is to note that each spatial point $\vec{r}$ where $\Re\sigma(\vec{r},\omega) > 0$ effectively represents an independent dissipative bath, uncorrelated with those at neighboring points. All subsequent steps are quite analogous to the circuit case, even though some short-distance regularization is required to make them well-defined mathematically. Obviously, $\sigma(\vec{r},\omega)$ of the material plays the same physical role as the admittance $1/Z_j(\omega)$ of circuit elements.

First of all, we define a function $\kappa(\vec{r},\omega)$, holomorphic for $\Im\omega > 0$ and such that $|\kappa(\vec{r},\omega)|^2 = \Re\sigma(\vec{r},\omega)$. For a Drude metal,
\begin{equation}
\sigma(\omega)=\frac{\sigma_\text{dc}}{1-i\omega\tau},\quad
\kappa(\omega)=\frac{\sqrt{\sigma_\text{dc}}}{1-i\omega\tau},
\end{equation}
while for an insulator with an optical phonon resonance, described by the Drude-Lorentz model,
\begin{subequations}\begin{align}
&\sigma(\omega)=\frac{\omega_L^2-\omega_T^2}{4\pi}\,\frac{i\omega}{\omega^2+2i\gamma\omega-\omega_T^2},\\
&\kappa(\omega)=\sqrt{\frac{\gamma(\omega_L^2-\omega_T^2)}{4\pi}}\,\,\frac{i\omega}{\omega^2+2i\gamma\omega-\omega_T^2}.
\end{align}\end{subequations}
Instead of the current and voltage at each resistor, we have the current density and the electric field at each point $\vec{r}$ with Cartesian components labeled by $k,l=x,y,z$:
\begin{widetext}
\begin{subequations}\begin{align}
\label{eq:current_density=}
&\hat{j}_k(\vec{r},t)=\int_0^\infty{d}\omega
\sqrt{\frac{\hbar\omega}{4\pi}}
\left[\frac{\sigma(\vec{r},\omega)}{\kappa(\vec{r},\omega)}\,\hat{a}_{k,\vec{r},\omega}^{+}+\frac{\sigma^*(\vec{r},\omega)}{\kappa^*(\vec{r},\omega)}\,\hat{a}_{k,\vec{r},\omega}^{-}\right]{e}^{-i\omega{t}} +\mbox{h.c.},\\
\label{eq:electric_field=}
&\hat{E}_k(\vec{r},t)=\int_0^\infty{d}\omega
\sqrt{\frac{\hbar\omega}{4\pi}}
\left[\frac{1}{\kappa(\vec{r},\omega)}\,\hat{a}_{k,\vec{r},\omega}^{+} -\frac{1}{\kappa^*(\vec{r},\omega)}\,\hat{a}_{k,\vec{r},\omega}^{-}\right]{e}^{-i\omega{t}} +\mbox{h.c.}.
\end{align}\end{subequations}
\end{widetext}
The bosonic operators $\hat{a}_{k,\vec{r},\omega}^{-}$ of the incident matter excitations have the commutation relations $[\hat{a}^{-}_{k,\vec{r},\omega},\hat{a}^{-\,\dagger}_{l,\vec{r}',\omega'}]=\delta_{kl}\,\delta(\vec{r}-\vec{r}')\,\delta(\omega-\omega')$, and thermal averages  $\langle \hat{a}^{-\,\dagger}_{k,\vec{r},\omega}\,\hat{a}^-_{l,\vec{r}',\omega'}\rangle=\delta_{kl}\,\delta(\vec{r}-\vec{r}')\, \delta(\omega-\omega')\,n(\vec{r},\omega)$, where $n(\vec{r},\omega)=[e^{-\hbar\omega/T(\vec{r})}-1]^{-1}$ is the Bose-Einstein distribution with the local temperature $T(\vec{r})$. The operators $\hat{a}_{k,\vec{r},\omega}^{+}$ of the scattered excitations are related to the incident ones via the scattering matrix, which now also involves integration over spatial variables:
\begin{equation}
\hat{a}_{k,\vec{r},\omega}^{+}=\sum_l\int{d}^3\vec{r}'\,\mathcal{S}_{kl}(\vec{r},\vec{r}')\,\hat{a}_{l,\vec{r}',\omega}^{-}.
\end{equation}
The scattering matrix can be found by solving Maxwell's equations in the given structure. To see how it works, let us represent the total current density as a sum of the fluctuating (quantum Langevin) source and the current induced by the electric field, which is conveniently done in the frequency domain: $\hat{\vec{j}}(\vec{r},\omega)=\sigma(\vec{r},\omega)\hat{\vec{E}}(\vec{r},\omega)+\hat{\vec{j}}^\text{fl}(\vec{r},\omega)$. Eq.~(\ref{eq:current_density=}) then gives
\begin{equation}
\hat{j}_k^\text{fl}(\vec{r},t)=\int_0^\infty{d}\omega
\sqrt{\frac{\hbar\omega}{4\pi}}
\,2\kappa(\vec{r},\omega)\,\hat{a}_{k,\vec{r},\omega}^{-}{e}^{-i\omega{t}} +\mbox{h.c.}.
\end{equation}
The electric field operator satisfies the Maxwell's equations with the total current as  a source. Separating the fluctuating part, we arrive at the familiar equation
\begin{align}\label{eq:Maxwell}
&\grad\times\grad\times\hat{\vec{E}}(\vec{r},\omega)
-\frac{\omega^2}{c^2}\left[1+\frac{4\pi\sigma(\vec{r},\omega)}{-i\omega}\right]\hat{\vec{E}}(\vec{r},\omega) 
\nonumber\\
{}&{} = \frac{4\pi{i}\omega}{c^2}\,\hat{\vec{j}}^\text{fl}(\vec{r},\omega),
\end{align}
which relates the electric field to the fluctuating current sources (the expression in the square brackets is nothing but the dielectric function of the medium). Its solution in a given structure can be written as
\begin{equation}
\hat{E}_k(\vec{r},\omega)=-\sum_l\int{d}^3\vec{r}'\,\frac{\mathcal{D}_{kl}^R(\vec{r},\vec{r}',\omega)}{-i\omega}\,\hat{j}_l^\text{fl}(\vec{r}',\omega),
\end{equation}
where $\mathcal{D}_{kl}^R(\vec{r},\vec{r}',\omega)$ is the retarded Green's function of the electric field, which we discuss in Appendix~\ref{App:PolarizationOp}.
Note that $\mathcal{D}_{kl}^R(\vec{r},\vec{r}',\omega)/(i\omega)$ plays the same role as the matrix $\zeta_{ij}(\omega)$ played for circuits (response of the voltages to external current sources). Matching Eq.~(\ref{eq:electric_field=}), we relate the scattering matrix to the electric field Green's function:
\begin{align}
\mathcal{S}_{kl}(\vec{r},\vec{r}') = {}&{} 2\kappa(\vec{r},\omega)\,\frac{\mathcal{D}^R_{kl}(\vec{r},\vec{r}',\omega)}{i\omega}\,\kappa(\vec{r}',\omega)\nonumber\\
{}&{} +{}
\delta_{kl}\,\delta(\vec{r}-\vec{r}')\,\frac{\kappa(\vec{r},\omega)}{\kappa^*(\vec{r},\omega)}.
\end{align}
The dissipated power density $\hat{\mathcal{P}}(\vec{r})$ at each point~$\vec{r}$ is given by the usual Joule losses:
\begin{widetext}\begin{subequations}\begin{align}
&\hat{\mathcal{P}}(\vec{r},t) =  \frac{1}2\left[\hat{\vec{E}}(\vec{r},t)\cdot\hat{\vec{j}}(\vec{r},t)+\hat{\vec{j}}(\vec{r},t)\cdot\hat{\vec{E}}(\vec{r},t)\right]\nonumber\\
{}&{} \qquad\quad = -\int_{-\infty}^\infty \frac{d \omega'\, d \omega''}{4 \pi}\, \hbar\sqrt{|\omega'\omega''|} \,e^{i (\omega'-\omega'')t}  \sum_{l',l''} \int{d^3}\vec{r}'\,d^3\vec{r}''\,\mathcal{A}_{l'l''}(\vec{r},\vec{r}',\vec{r}'';\omega',\omega'') \, \hat\alpha_{l',\vec{r}',\omega'}^\dagger\hat\alpha_{l'',\vec{r}'',\omega''},
\end{align}
with $\hat\alpha_{k,\vec{r},\omega>0}\equiv\hat{a}_{k,\vec{r},\omega}^-$, $\hat\alpha_{k,\vec{r},\omega<0}\equiv\hat{a}_{k,\vec{r},\omega}^{-\,\dagger}$, and
\begin{align}
\mathcal{A}_{l'l''}(\vec{r},\vec{r}',\vec{r}'';\omega',\omega'') =&
-\frac{\sigma^*(\vec{r},\omega')+\sigma(\vec{r},\omega'')}{2\kappa^*(\vec{r},\omega')\,\kappa(\vec{r},\omega'')}
\sum_k\mathcal{S}^*_{kl'}(\vec{r},\vec{r}',\omega')\,\mathcal{S}_{kl''}(\vec{r},\vec{r}'',\omega'')
\nonumber\\
\hspace*{3.4cm} {}&-\frac{\sigma(\vec{r},\omega')-\sigma(\vec{r},\omega'')}{2\kappa(\vec{r},\omega')\,\kappa(\vec{r},\omega'')}\,
\delta(\vec{r}-\vec{r}')\,\mathcal{S}_{l'l''}(\vec{r},\vec{r}'',\omega'')
+\frac{\sigma^*(\vec{r},\omega')-\sigma^*(\vec{r},\omega'')}{2\kappa^*(\vec{r},\omega')\,\kappa^*(\vec{r},\omega'')}\,\mathcal{S}_{l''l'}^*(\vec{r},\vec{r}',\omega')\,\delta(\vec{r}-\vec{r}'')\nonumber\\
\hspace*{3.4cm} {}
&+\frac{\sigma(\vec{r},\omega')+\sigma^*(\vec{r},\omega'')}{2\kappa(\vec{r},\omega')\,\kappa^*(\vec{r},\omega'')}\,\delta_{l'l''}\,\delta(\vec{r}-\vec{r}')\,\delta(\vec{r}-\vec{r}''),
\end{align}
which equals
\begin{align}
\mathcal{A}_{l'l''}(\vec{r},\vec{r}',\vec{r}'';\omega',\omega'')=& -2\,\kappa^*(\vec{r}',\omega')\,\kappa(\vec{r}'',\omega'')[\sigma^*(\vec{r},\omega')+\sigma(\vec{r},\omega'')]
\sum_k\frac{\mathcal{D}_{kl}^{R*}(\vec{r},\vec{r}',\omega')\,\mathcal{D}^R_{kl''}(\vec{r},\vec{r}'',\omega'')}{\omega'\omega''}\nonumber\\
\hspace*{3.4cm} {}
&-2\,\kappa^*(\vec{r}',\omega')\,\kappa(\vec{r}'',\omega'')
\left[\delta(\vec{r}-\vec{r}')\,\frac{\mathcal{D}^R_{l'l''}(\vec{r},\vec{r}'',\omega'')}{-i\omega''}
+\frac{\mathcal{D}_{l''l'}^{R*}(\vec{r},\vec{r}',\omega')}{i\omega'}\,\delta(\vec{r}-\vec{r}'')\right].
\end{align}\end{subequations}\end{widetext}
Then, one can calculate arbitrary correlators of the power density, like $\langle\hat{\mathcal{P}}(\vec{r},t)\,\hat{\mathcal{P}}(\vec{r}',t')\rangle$, etc.

The scattering matrix $\mathcal{S}_{kl}(\vec{r},\vec{r}',\omega)$, even though a natural object in any scattering problem, in the present setting may lead to mathematically ill-defined expressions, which should be handled with care. For example, in the average dissipated power we have to deal with the square of the $\delta$ function,
\begin{align}
\langle\hat{\cal P}(\vec{r})\rangle = {}&{}
\int_0^\infty\frac{d\omega}{4\pi}\,\hbar\omega
\int{d^3}\vec{r}'\,[2n(\vec{r}',\omega)+1]\nonumber\\
{}&{}\times\left[\sum_{k,l}|\mathcal{S}_{kl}(\vec{r},\vec{r}',\omega')|^2-3\delta^2(\vec{r}-\vec{r}')\right]\nonumber\\
= {}&{} \int_0^\infty\frac{d\omega}{2\pi}\,\hbar\omega
\int{d^3}\vec{r}'\,\sum_{k,l}|\mathcal{S}_{kl}(\vec{r},\vec{r}',\omega')|^2\nonumber\\
{}&{}\times[n(\vec{r}',\omega)-n(\vec{r},\omega)]
\end{align}
where we used $n(\vec{r}',\omega)\,\delta^2(\vec{r}-\vec{r}')=n(\vec{r},\omega)\,\delta^2(\vec{r}-\vec{r}')$, as well as the unitarity of the scattering matrix in the form
\begin{equation}
\sum_l\int{d}^3\vec{r}'\,|\mathcal{S}_{kl}(\vec{r},\vec{r}',\omega')|^2
=\delta(\vec{r}=0)
=\int{d}^3\vec{r}'\,\delta^2(\vec{r}-\vec{r}').
\end{equation}

From the practical point of view, it is more convenient to equivalently express the result in terms of $\mathcal{D}_{kl}^R(\vec{r},\vec{r}',\omega)$, which does not suffer from short-distance singularities:
\begin{align}
\langle\hat{\cal P}(\vec{r})\rangle = {}&{}
\int\limits_0^\infty\frac{d\omega}{\pi}\,\hbar\omega\sum_{k,l}
\int{d^3}\vec{r}'\,[2n(\vec{r}',\omega)+1]\,\frac{\Re\sigma(\vec{r},\omega)}\omega\nonumber\\
{}&{}\times\left[|\mathcal{D}_{kl}^R(\vec{r},\vec{r}',\omega')|^2\,\frac{\Re\sigma(\vec{r}',\omega)}\omega\right.\nonumber\\
{}&{}\quad{}+\left.\Im\mathcal{D}_{kk}^R(\vec{r},\vec{r},\omega)\,\delta_{kl}\,\delta(\vec{r}-\vec{r}')\right].
\end{align}
Using the relation
\begin{equation}
\Im\mathcal{D}^R_{kk}(\vec{r},\vec{r},\omega) = -\sum_l\int{d}^3\vec{r}'\,|\mathcal{D}_{kl}^R(\vec{r},\vec{r}',\omega')|^2\,\frac{\Re\sigma(\vec{r}',\omega)}\omega,
\end{equation}
which can be obtained from Eq.~(\ref{eq:Maxwell}) and its complex conjugate,
we arrive at the standard Caroli formula for the dissipated power:
\begin{align}
\langle\hat{\cal P}(\vec{r})\rangle = {}&{}
\int\limits_0^\infty\frac{d\omega}{\pi}\,\hbar\omega\sum_{k,l}
\int{d^3}\vec{r}'\,[n(\vec{r}',\omega)-n(\vec{r},\omega)]\nonumber\\
{}&{}\times 2\,\frac{\Re\sigma(\vec{r},\omega)}\omega\,|\mathcal{D}_{kl}^R(\vec{r},\vec{r}',\omega')|^2\,\frac{\Re\sigma(\vec{r}',\omega)}\omega.
\end{align}
Note that $\sigma(\vec{r},\omega)/(-i\omega)$ is nothing but the retarded polarization operator of the medium, see Eq.~(\ref{eq:PiR=}). In the same way, one can obtain equivalent expressions for higher-order correlators of dissipated power in terms of the scattering matrix or in terms of the matter and field Green's functions.

The presented scattering problem construction is fully analogous to that in circuits. The situation becomes a bit less straightforward if the spatial dispersion should be taken into account in the optical conductivity. In this case, it becomes an integral operator with the kernel $\sigma_{kl}(\vec{r},\vec{r}',\omega)$, and one cannot treat each spatial point as an independent bath. Instead, at each $\omega$ one should diagonalize the operator $\Re\sigma_{kl}(\vec{r},\vec{r}',\omega)$, and each eigenmode of this operator can be represented as an independent scattering channel with incident and scattered excitation. Then the construction again becomes analogous, although solution of the Maxwell's equations becomes quite difficult, as is usually the case in the presence of spatial dispersion.

Another complication arises from the fact that besides the absorbing bodies, the system may have one more dissipative bath, corresponding to radiation at $|\vec{r}|\to\infty$. This depends on the geometry of the problem: while for the textbook example of heat transfer between two semi-infinite half-spaces (metallic or dielectric), separated by a vacuum gap, all radiation is eventually absorbed by the material, the case of two spheres involves photons that can escape to infinity, as well as thermal photons incident from infinity. These represent additional scattering channels, and only upon inclusion of these channels the scattering matrix becomes unitary. More detailed study of this issue is beyond the scope of our work.

\section{Conclusions}
\label{sec:Conclusion}

We have developed a quantum-mechanical scattering theory for NFRHT problem, which enables one to calculate fluctuations of the energy current. Our general expression differs from the B\"uttiker's formula for particle current fluctuations in the presence of reactive components in the energy reservoir. We show that this reactive contribution can dominate the energy current fluctuations at finite frequencies. To determine it correctly, it is important to precisely define how the energy current is measured.

Our construction describes scattering of matter excitations due to their coupling to the electromagnetic field, and has some limitations. Temperature dependence of the scattering matrix is not naturally included, similarly to the Landauer-B\"uttiker approach to electron transport; temperature dependence arises from many-body effects inside the material (such as electron-electron interactions or phonon anharmonicity), so a description in terms of scattering of single excitations is, at best, approximate, if valid at all. Heat transfer via phonons or electrons is far beyond our construction (in the same way as it is beyond FED), since it represents a totally different physical mechanism involving matter excitations themselves.

\begin{acknowledgments}
This research was supported by the Deutsche Forschungsgemeinschaft (DFG; German Research Foundation) via SFB 1432 (Project \mbox{No.} 425217212).
\end{acknowledgments}

\appendix

\section{Resistor under a classical drive}
\label{App:ResistorClassicalDrive}

Here we show that Eqs.~\eqref{eq:IV=modes} for the voltage and current operators are consistent with the standard definition of the impedance.
The impedance $Z(\omega)$ of a circuit element can be defined either via the response of the  current through this element to an externally applied monochromatic voltage source, or via the response of the voltage to an external current source.

Let us connect an external voltage source $V_\text{ext}e^{-i\omega_\text{ext}t}+V_\text{ext}^*e^{i\omega_\text{ext}t}$ to the $j$th resistor. This means that the voltage operator $\hat{V}_j(t)$ is constrained to have this value, thus becoming a pure number. Interpreting this condition,
\begin{align}
&\int_0^\infty{d}\omega
\sqrt{\frac{\hbar\omega}{4\pi}}\left[ \frac{Z_j(\omega)}{\lambda_j(\omega)}\,\hat{a}_{j,\omega}^++\frac{Z_j^*(\omega)}{\lambda_j^*(\omega)}\,\hat{a}_{j,\omega}^- \right]{e}^{-i\omega{t}}
+\mbox{h.c.}\nonumber\\
{}&{}= V_\text{ext}e^{-i\omega_\text{ext}t}+\mbox{c.c.},
\end{align}
as an equation for $\hat{a}_{j,\omega}^+$, we apply Fourier transform on both sides, and find
\begin{equation}\label{eq:voltage_driven_aplus}
    \hat{a}_{j,\omega}^+ = V_\text{ext}\,\frac{\lambda_j(\omega)}{Z_j(\omega)}\sqrt{\frac{4\pi}{\hbar\omega}}\,\delta(\omega-\omega_\text{ext}) - \frac{Z_j^*(\omega)}{Z_j(\omega)}\,\frac{\lambda_j(\omega)}{\lambda_j^*(\omega)}\,\hat{a}_{j,\omega}^-.
\end{equation}
Such linear shift of the bosonic operators in the Heisenberg picture implies for the Schr\"odinger picture that if the incident excitations were in the vacuum state, then the scattered excitations are in the coherent state, resulting in a non-zero expectation value of the current. This expectation value can be easily found by averaging the Heisenberg operator $\hat{I}_j(t)$ [Eq.~\eqref{eq:It=modes}] over the vacuum state. Only the first term in Eq.~(\ref{eq:voltage_driven_aplus}) contributes to this non-zero average, yielding
\begin{equation}
    \langle\hat{I}_j(t)\rangle = \frac{V_\text{ext}}{Z_j(\omega_\text{ext})}\,e^{-i\omega_\text{ext}{t}} + \mbox{c.c.},
\end{equation}
so that the impedance is indeed $Z_j(\omega)$.
The same result is obtained if the incident excitations are in a thermal state. 

If we connect an external current source $I_\text{ext}e^{-i\omega_\text{ext}t} + I_\text{ext}^*e^{i\omega_\text{ext}t}$, this results in imposing
\begin{align}
    &\int_0^\infty{d}\omega
\sqrt{\frac{\hbar\omega}{4\pi}}
\left[\frac{1}{\lambda_j(\omega)}\,\hat{a}_{j,\omega}^{+}-\frac{1}{\lambda^*_j(\omega)}\,\hat{a}_{j,\omega}^{-}\right]{e}^{-i\omega{t}} +\mbox{h.c.} \nonumber\\
{}&{} = I_\text{ext}e^{-i\omega_\text{ext}t} + \mbox{c.c.}.
\end{align}
Again, this condition enables us to express the scattered excitation operator as
\begin{equation}\label{eq:current_driven_aplus}
    \hat{a}_{j,\omega}^+ = I_\text{ext}\lambda_j(\omega)\sqrt{\frac{4\pi}{\hbar\omega}}\,\delta(\omega-\omega_\text{ext}) + \frac{\lambda_j(\omega)}{\lambda^*_j(\omega)}\,\hat{a}_{j,\omega}^-,
\end{equation}
yielding the average voltage
\begin{equation}
    \langle\hat{V}_j(t)\rangle = I_\text{ext}Z_j(\omega_\text{ext})\,e^{-i\omega_\text{ext}{t}} + \mbox{c.c.},
\end{equation}
again corresponding to the impedance $Z_j(\omega)$.

In this picture, the Joule losses in the resistor correspond to emission of excitations into the transmission line. This can be checked by employing Eq.~(\ref{eq:voltage_driven_aplus}) or Eq.~(\ref{eq:current_driven_aplus}) to calculate the average of the power $\langle\hat{P}_j(t)\rangle$.

\section{Simple circuits}
\label{App:SimpleCirc}
Here we discuss simple circuits to illustrate the impact of reactive elements within the thermal reservoir, which becomes evident in the heat-current noise spectrum but not in the average heat current. We calculate the noise spectrum for the specific examples of a low-pass and resonant-pass filter. Both cases can be viewed as two impedances $Z_{1,2}(\omega)$, galvanically connected in a loop. The scattering matrix for this configuration is straightforwardly found from Eqs.~\eqref{eq:IV=modes} and the conditions $\hat{V}_1=\hat{V}_2$, $\hat{I}_1=-\hat{I}_2$: 
\begin{equation}
    S = \frac{1}{Z_1+Z_2}\begin{pmatrix}
        \lambda_1(Z_2-Z_1^*)/\lambda_1^* & 2\lambda_1 \lambda_2 \\ 
        2 \lambda_1 \lambda_2 & \lambda_2(Z_1-Z_2^*)/\lambda_2^*
    \end{pmatrix}.
\end{equation}
We focus on the power flowing into reservoir 1 and calculate $W_{11}(\Omega)$ assuming for simplicity $T_1\ll{T}_2$. The B\"uttiker-like contribution $W^B_{11}(\Omega)$ matches with the total heat-current noise spectrum $W_{11}(\Omega)$ when the reactive elements are excluded from the reservoir and included in the coupling circuit. In these specific examples, the heat-current noise spectra $W_{11}(\Omega)$ and $W_{22}(\Omega)$ coincide, since the current and voltage across the impedance $Z_1(\omega)$ are the same as those across the resistor $R$.

According to Eq.~\eqref{eq:Wjk=}, $W_{11}(\Omega)$ is given by a sum of four terms, $\sum_{l,m}W_{11}^{(lm)}(\Omega)$, each containing a factor $\coth[\hbar\omega/(2T_l)]\coth[\hbar(\omega+\Omega)/(2T_m)]-1$ in the integrand for $l,m=1,2$.  The term $W^{(22)}_{11}(\Omega)$ is due to excitations originating in the reservoir with temperature~$T_2$, $W^{(11)}_{11}(\Omega)$ is due to excitations from the reservoir with temperature $T_1 \ll T_2$, and thus is negligibly small. The terms $W^{(12)}_{11}(\Omega)$ and $W^{(21)}_{11}(\Omega)$ are due to interference.

\begin{figure}
    \centering
    \includegraphics[width=\columnwidth]{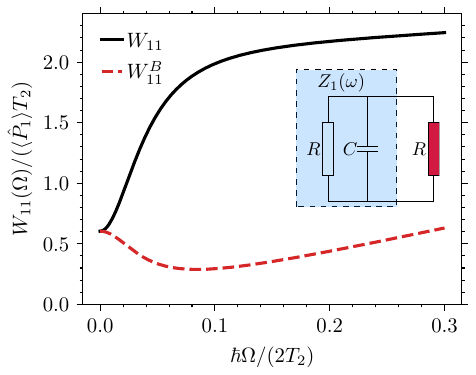}
    \caption{Heat-current noise spectrum $W_{11}(\Omega)$ (solid line) and the B\"uttiker-like contribution $W_{11}^B(\Omega)$ (dashed line) for the circuit shown in the inset. The impedance $Z_1(\omega)$ consists of a parallel $RC$ element galvanically coupled to the second resistor $R$. Here, the $RC$ time is chosen as $T_2 RC/\hbar=50$. Impedance temperature $T_1$ is assumed to be much smaller than the temperature $T_2$ of the second resistor.  }
\label{fig:low-pass}
\end{figure}

\begin{figure}
\centering
    \includegraphics[width=\columnwidth]{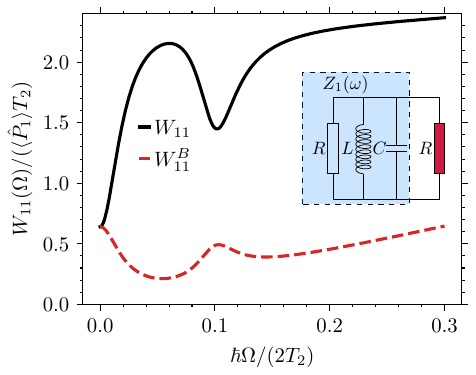}
    \caption{Heat-current noise spectrum $W_{11}(\Omega)$ (solid line) and the B\"uttiker-like contribution $W_{11}^B(\Omega)$ (dashed line) for the circuit shown in the inset. The impedance $Z_1(\omega)$ consists of a parallel $RLC$ element galvanically coupled to the second resistor $R$. The $RC$ time is chosen as $ T_2 RC/\hbar=50$ and the resonance frequency as $ T_2 \sqrt{LC} /\hbar=10$. Impedance temperature $T_1$ is assumed to be much smaller than the temperature $T_2$ of the second resistor. }
\label{fig:resonant-pass}
\end{figure}

The low-pass filter consists of a parallel $RC$ element with the impedance $Z_1(\omega)=R/(1-i\omega{R}C)$, coupled galvanically to a second resistor $Z_2(\omega)=R$ (see Fig.~\ref{fig:low-pass}). Such circuit may represent an electronic tunnel junction. Choosing $\lambda_1(\omega)=\sqrt{R}/(1-i\omega{R}C)$, $\lambda_2(\omega)=\sqrt{R}$, we find the transmission function given by a Lorentzian
\begin{equation}
|S_{12}(\omega)|^2=\frac1{1+\omega^2R^2C^2/4},
\end{equation}
where the $RC$-time corresponds to the inverse of the damping parameter. For $\hbar/(RC)\ll T_2$, the transmission function is sharply peaked around $\omega=0$, and the average power is approximately $\langle \hat{P}_1 \rangle \approx T_2/(2RC)$. 

Assuming $\hbar\Omega\ll{T}_2$, we evaluate the frequency integral in Eq.~\eqref{eq:Wjk=}, and find the B\"uttiker-like contribution given by
\begin{equation}
W^{B(22)}_{11}(\Omega) \approx 
\frac{T_2^2}{4RC}\,\frac{1}{1+(\Omega{R}C/4)^2}.
\end{equation}
The full $W^{(22)}_{11}(\Omega)$ contains an additional factor
\[
\frac{|Z_1(\omega)+Z_1^*(\omega+\Omega)|^2}{4 \Re Z_1(\omega) \Re Z_1(\omega+\Omega)}=1+(\Omega{R}C/2)^2,
\]
which leads to
\begin{equation}
W^{(22)}_{11}(\Omega) \approx 
\frac{T_2^2}{4RC}\,\frac{1+(\Omega{R}C/2)^2}{1+(\Omega{R}C/4)^2}.
\end{equation}
The two expressions behave quite differently; the B\"uttiker-like contribution peaks at $\Omega=0$ and decays for $\Omega RC \gg 1$, while $W^{(22)}_{11}(\Omega)$ exhibits a dip at $\Omega=0$.
The interference terms $W^{B(12)}_{11}(\Omega)=W^{B(21)}_{11}(\Omega)$ are subleading in the limit $T_2\gg\hbar\Omega$, but they grow linearly with frequency, $W^{B(12)}_{11}(\Omega)\approx T_2 \hbar |\Omega|/(4RC)$, in the limit $\Omega RC \gg 1$.
The interference terms play a minor role in the total spectrum $W_{11}(\Omega)$, where $W^{(22)}_{11}(\Omega)$ dominates the behavior.

In the second example, two resistors are coupled via an $LC$ resonator (see Fig.~\ref{fig:resonant-pass}), which may model a small metallic particle hosting a plasmonic resonance (represented by the $LC$ resonator) together with electron-hole pair excitations (the resistor). The circuit partitioning is then determined by the energy measurement protocol, which may be either sensitive to electron-hole pairs only, or to the plasmonic energy as well. Let us calculate $W_{11}(\Omega)$ assuming that the $LC$ resonator is included in the cold reservoir whose impedance is thus
\begin{equation}
    Z_1(\omega) = \left(\frac{1}R-\frac{1}{i\omega{L}}-i\omega{C}\right)^{-1},
\end{equation}
then the B\"uttiker-like contribution $W_{11}^B(\Omega)$ will correspond to the first reservoir including only the resistor.
We can choose
\begin{equation}
    \lambda_1(\omega)=\frac{-i\omega L/\sqrt{R}}{1-\omega^2 LC-i\omega L/R},
\end{equation}
which results in the transmission function
\begin{equation}
|S_{12}(\omega)|^2=\frac{(2\omega\gamma)^2}{(\omega^2-\omega_0^2)^2 +(2\omega\gamma)^2},
\end{equation}
where we denoted $\omega_0\equiv1/\sqrt{LC}$, $\gamma\equiv1/(RC)$.
The transmission function exhibits peaks at $\omega=\pm \omega_0$ resulting in peaks at $\Omega=0,\pm 2\omega_0$ in the B\"uttiker component $W^{B(22)}_{11}(\Omega)$~\cite{Wise:2022} Again, the increase of $W^{B}_{11}(\Omega)$ for $\Omega \gtrsim 2\omega_0$ stems from the interference terms $W^{B(12)}_{11}(\Omega)=W^{B(21)}_{11}(\Omega)$. The full noise spectrum $W_{11}(\Omega)$ is quite different from $W_{11}^B(\Omega)$. Instead of peaks, $W_{11}(\Omega)$ displays dips at the locations $\Omega=0,\pm 2 \omega_0$. Its shape is primarily determined by $W^{(22)}_{11}(\Omega)$ and not by the interference terms $W^{(12)}_{11}(\Omega)=W^{(21)}_{11}(\Omega)$.

\section{Effective circuit for coupled surface polaritons}
\label{sec:circuit_polaritons}

Two coupled surface polariton modes can arise in the planar geometry in the TM polarization when two semi-infinite half-spaces with dielectric function $\varepsilon(\omega)$ are separated by a vacuum gap of width~$d$. In the quasi-static limit, corresponding to the speed of light $c\to\infty$, the dispersion $\omega_\pm(q)$ of the symmetric/antisymmetric mode as a function of the in-plane wave vector~$q$ is determined by the equation 
\begin{equation}
\varepsilon(\omega) = -\frac{1\mp{e}^{-qd}}{1\pm{e}^{-qd}}.
\end{equation}
For the Drude-Lorentz model of optical phonon resonance in a dielectric,
\begin{equation}\label{eq:Drude-Lorentz-nogamma}
\varepsilon(\omega) = \varepsilon_\infty\,\frac{\omega^2-\omega_L^2}{\omega^2-\omega_T^2},
\end{equation}
the frequencies of the symmetric and antisymmetric modes are given by
\begin{equation}\label{eq:wpmq=}
\omega^2_\pm(q) = \frac{\varepsilon_\infty\omega_L^2+\omega_T^2\pm (\varepsilon_\infty\omega_L^2-\omega_T^2)e^{-qd}}{\varepsilon_\infty+1\pm (\varepsilon_\infty-1)e^{-qd}}.
\end{equation}
Here $\omega_T$ is the mechanical frequency of the triply degenerate polar optical phonon, and $\omega_L$~is the Coulomb frequency of the longitudinal optical phonon (both in the bulk). 
A finite damping~$\gamma$ of the optical phonon can be included by replacing $\omega^2\to\omega^2+2i\gamma\omega$ in Eq.~(\ref{eq:Drude-Lorentz-nogamma}).
At $d\to\infty$ the splitting vanishes, and the two surfaces are decoupled, each hosting its own weakly damped surface polariton. 

\begin{figure}
\begin{center}
\includegraphics[width=0.3\textwidth]{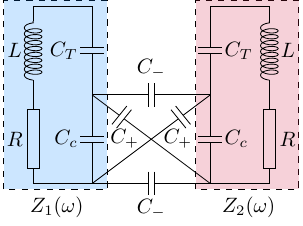}
\end{center}
\caption{The model circuit qualitatively mimicking heat transfer via surface polaritons.}
\label{fig:polariton_circuit}
\end{figure}

This situation can be modeled by the circuit in Fig.~\ref{fig:polariton_circuit}.
Here each $RLC$ contour hosts a weakly damped resonance at frequency $\omega_L=\sqrt{\omega_T^2+1/(LC_\text{c})}$, where we denoted $\omega_T^2\equiv1/(LC_T)$.
The two contours are coupled by capacitances~$C_\pm$. The frequency of the coupled symmetric/antisymmetric mode is determined by the equation
\begin{equation}\label{eq:omegapm_eq}
-i\omega{C}_\text{c} - \frac{i\omega{C}_T}{1-\omega^2LC_T-i\omega{R}C_T} = i\omega{C}_\pm.
\end{equation}
We are interested in the regime when the damping is smaller than the splitting, so that the doublet is well resolved.
Then the solutions for the frequencies are approximately given by
\begin{equation}
    \omega_\pm-i\gamma\equiv\sqrt{\frac{1}{LC_T}+\frac{1}{L(C_\text{c}+C_\pm)}}-i\,\frac{R}{2L},
\end{equation}
where $\omega_\pm$ are the hybridized modes' frequencies, and $\gamma$~is their amplitude decay rate. For the condition $\gamma\ll|\omega_+-\omega_-|$ to hold, the circuit parameters should satisfy
\begin{equation}
    R\ll\frac{|C_+-C_-|}{\omega_L(C_\text{c}+C_+)(C_\text{c}+C_-)}.
\end{equation}


The Poynting vector in the middle of the vacuum gap between the two dielectrics corresponds to the power deposited into the $RLC$ contour, which is thus treated as a dissipative bath with impedance
\begin{align}
Z(\omega) = {}&{} \left[-i\omega{C}_\text{c} - \frac{i\omega{C}_T}{1-\omega^2LC_T-i\omega{R}C_T}\right]^{-1}
\nonumber\\
= {}&{}
\frac{1}{-i\omega{C}_\text{c}}\,\frac{\omega_T^2-\omega^2-2i\omega\gamma}{\omega_L^2-\omega^2-2i\omega\gamma},
\end{align}
while the capacitances $C_\pm$ represent the coupling circuit.
According to the general construction of the main text, this association leads to
\begin{align}
&\Re{Z}(\omega) = \frac{R/(LC_\text{c})^2}{(\omega_L^2-\omega^2)^2+(2\omega\gamma)^2},\\
&\lambda(\omega) = \frac{\sqrt{R}/(LC_\text{c})}{\omega_L^2-\omega^2-2i\omega\gamma},
\end{align}
and
\begin{subequations}\begin{align}
&\zeta_{11}=\zeta_{22} = \frac{2/Z-i\omega(C_++C_-)}{2(1/Z-i\omega{C}_+)(1/Z-i\omega{C}_-)},\\
&\zeta_{12}=\zeta_{21} = \frac{i\omega(C_+-C_-)}{2(1/Z-i\omega{C}_+)(1/Z-i\omega{C}_-)}.
\end{align}\end{subequations}
According to Eq.~\eqref{eq:S}, the scattering matrix is given by
\begin{subequations}\label{eqs:SmatrixPolariton=}\begin{align}
&S_{11}=S_{22}=\frac{\lambda}{\lambda^*}\frac{1+\omega (C_++C_-) \Im{Z}+\omega^2  C_+ C_- |Z|^2 }{(1-i \omega C_+Z)(1-i \omega C_-Z)},\\
&S_{12}=S_{21}=\frac{i \omega \lambda^2 (C_+-C_-) }{(1-i \omega C_+Z)(1-i \omega C_-Z)},
\end{align}\end{subequations}
leading to the transmission and reflection coefficients
\begin{subequations}\begin{align}
    |S_{12}(\omega)|^2= {}&{} \frac{[\omega(C_+-C_-)\Re{Z}(\omega)]^2}{|1-i \omega C_+Z(\omega)|^2|1-i \omega C_-Z(\omega)|^2}\nonumber\\
    ={}&{}\frac{(2\gamma\omega)^2(\omega_+^2-\omega_-^2)^2}{[(\omega_+^2-\omega^2)^2+(2\gamma\omega)^2][(\omega_-^2-\omega^2)^2+(2\gamma\omega)^2]},
    \label{eq:S21modsquare}\\
    |S_{11}(\omega)|^2= {}&{} 1-|S_{12}(\omega)|^2\nonumber\\
    ={}&{}\frac{[(\omega_+^2-\omega^2)(\omega_-^2-\omega^2)+(2\gamma\omega)^2]^2}{[(\omega_+^2-\omega^2)^2+(2\gamma\omega)^2][(\omega_-^2-\omega^2)^2+(2\gamma\omega)^2]}.
\end{align}    \end{subequations}
Due to the condition $|\omega_+-\omega_-|\gg\gamma$, we can approximating $|S_{12}(\omega)|^2$ by a sum of four Lorentzians,
\begin{equation}\label{eq:S12Lorentz}
    |S_{12}(\omega)|^2\approx\sum_{\sigma,\sigma'=\pm}\frac{\gamma^2}{(\omega+\sigma\omega_{\sigma'})^2+\gamma^2}.
\end{equation}
In this approximation, the average power is given by
\begin{equation}
    \langle\hat{\cal P}_1\rangle = \sum_{\sigma=\pm}\frac{\gamma}2\left(\frac{\hbar\omega_\sigma}{e^{\hbar\omega_\sigma/T_2}-1}-\frac{\hbar\omega_\sigma}{e^{\hbar\omega_\sigma/T_1}-1}\right).
\end{equation}
This expression can also be obtained from the simple rate equations for the bosonic occupation of the coupled resonant modes.

We are interested in the correlator $W_{11}(\Omega)$ of the power dissipated into the impedance~1 in the limit $T_2\gg{T}_1,\hbar\Omega$. According to Eq.~\eqref{eq:Wjk=}, $W_{11}(\Omega)$ is given by a sum of four terms, $\sum_{l,m}W_{11}^{(lm)}(\Omega)$, each containing a factor $\coth[\hbar\omega/(2T_l)]\coth[\hbar(\omega+\Omega)/(2T_m)]-1$ in the integrand for $l,m=1,2$. Let us analyze these contributions one by one.

For the contribution with $l=m=2$ we use the identity $\coth{x}\coth(x+y)-1=\cosh{y}/[\sinh{x}\sinh(x+y)]$ and write it as
\begin{align}
    W_{11}^{(22)}(\Omega) = {}&{} \cosh\frac{\hbar\Omega}{2T_2}
    \int_{-\infty}^\infty\frac{d\omega}{16\pi}\,
    \frac{\hbar\omega|S_{12}(\omega)|^2}{\sinh[\hbar\omega/(2T_2)]}\nonumber\\
    {}&{}\hspace*{1.5cm}\times\frac{\hbar(\omega+\Omega)|S_{12}(\omega+\Omega)|^2}{\sinh[\hbar(\omega+\Omega)/(2T_2)]}\nonumber\\
    {}&{}\hspace*{1.5cm}\times
    \frac{|Z(\omega)+Z^*(\omega+\Omega)|^2}{4\Re{Z}(\omega)\Re{Z}(\omega+\Omega)}.
    \label{eq:W11=}
\end{align}
The B\"uttiker part $W_{11}^{B(22)}(\Omega)$ is given by the same expression, but without the last line. Since $|S_{12}(\omega)|^2$ is peaked around $\omega=\omega_\pm,-\omega_\pm$,
the frequency integral in Eq.~(\ref{eq:W11=}) can be particularly large for such~$\Omega$ that the peaks in the two factors $|S_{12}(\omega)|^2$ and $|S_{12}(\omega+\Omega)|^2$ overlap. This happens near $\Omega=0,\pm|\omega_+-\omega_-|$. 

Let us first calculate the B\"uttiker contribution $W_{11}^{B(22)}(\Omega)$, which is simpler.
Using the Lorentzian approximation~(\ref{eq:S12Lorentz}) together with the fact that the convolution of two Lorentzians is a Lorentzian with a double width, and assuming $T_2\sim\hbar\omega_\pm\gg\hbar|\omega_+-\omega_-|$, we find
\begin{align}
W_{11}^{B(22)}(\Omega) = {}&{} \left[\frac{\hbar\omega_\pm/2}{\sinh[\hbar\omega_\pm/(2T_2)]}\right]^2\nonumber\\
{}&{}\times\left[\frac{2\gamma^3}{\Omega^2+4\gamma^2}+\frac{\gamma^3}{(\Omega+\omega_+-\omega_-)^2+4\gamma^2}\right.\nonumber\\
{}&{}\;\;\;+\left.\frac{\gamma^3}{(\Omega+\omega_--\omega_+)^2+4\gamma^2}\right],
\end{align}
where $\hbar\omega_\pm$ in the first factor means that either of the two frequencies can be taken within the made approximation.
To evaluate the full spectrum, $W_{11}^{(22)}(\Omega)$, we rewrite the last factor in Eq.~(\ref{eq:W11=}) as
\begin{align}
    \frac{|Z(\omega)+Z^*(\omega+\Omega)|^2}{4\Re{Z}(\omega)\Re{Z}(\omega+\Omega)}
    {}&{}= 1 + \frac{|Z(\omega+\Omega)-Z(\omega)|^2}{4\Re{Z}(\omega+\Omega)\Re{Z}(\omega)}\nonumber\\
    {}&{}\approx 1 + \frac{\Omega^2}{4\gamma^2},
\end{align}
where the last line is written in the approximation $|\omega-\omega_L|,\Omega,\gamma\ll(LC_\text{c})^{-1/2},(LC_T)^{-1/2}$. Then, we have a very simple relation
\begin{equation}
    W_{11}^{(22)}(\Omega) = \frac{\Omega^2+4\gamma^2}{4\gamma^2}\,W_{11}^{B(22)}(\Omega),
\end{equation}
which shows a rather dramatic effect of the reactive contribution: the central peak around $\Omega=0$ is fully suppressed, while the two peaks at $\Omega=\pm(\omega_+-\omega_-)$ are strongly enhanced.

Let us now consider the contribution $W_{11}^{(12)}(\Omega)$ with $l=1$ and $m=2$ [the other one is just $W_{11}^{(21)}=W_{11}^{(12)}(-\Omega)$]. The B\"uttiker contribution $W_{11}^{B(12)}(\Omega)$ is calculated quite analogously to $W_{11}^{B(22)}(\Omega)$, in the limit $T_1,\hbar\Omega\ll{T}_2$ and the Lorentzian approximation~(\ref{eq:S12Lorentz}):
\begin{align}
    W_{11}^{B(12)}(\Omega) = {}&{} 
    \int\limits_{-\infty}^\infty\frac{d\omega}{8\pi}\,
    \frac{\hbar^2\omega(\omega+\Omega)|S_{11}(\omega)|^2|S_{12}(\omega+\Omega)|^2}{e^{\sgn(\omega) \hbar(\omega+\Omega)/T_2}-1} \nonumber \\
     = {}&{} \frac{\hbar^2\omega_\pm^2/2}{e^{\hbar\omega_\pm/T_2}-1}\nonumber\\
{}&{}\times\left[\frac{\Omega^2\gamma+2\gamma^3}{\Omega^2+4\gamma^2}-\frac{\gamma^3}{(\Omega+\omega_+-\omega_-)^2+4\gamma^2}\right.\nonumber\\
{}&{}\;\;\;-\left.\frac{\gamma^3}{(\Omega+\omega_--\omega_+)^2+4\gamma^2}\right]. 
\end{align}
For the full spectrum $W_{11}^{(12)}(\Omega)$, we cannot give a simple compact analytical expression because of interference terms. Still, at $\Omega=0$ it obviously coincides with $W_{11}^{B(12)}(\Omega=0)$, while at $\Omega\gg\gamma$ it differs from $W_{11}^{B(12)}(\Omega)$ by a large factor, of the same origin as in $W_{11}^{(22)}(\Omega)$.

Finally, the contribution $l=m=1$ is negligibly small. Indeed, the factor $\coth[\hbar\omega/(2T_1)]\coth[\hbar(\omega+\Omega)/(2T_1)]-1$ is non-vanishing only in a narrow region $|\omega|\sim\max\{\Omega,T_1/\hbar\}$, so it misses the peaks at $|\omega|\approx\omega_\pm$. The rest of the integrand is regular at low frequencies, so the whole integral is small.

\section{Non-equilibrium Green's functions for coupled surface polaritons}
\label{App:GreenSurfacePolariton}

Here we study how the results for heat-current noise obtained from the circuit scattering approach are related to those found using non-equilibrium Green's functions. This latter approach is always based on a specific microscopic model for the involved materials, so we choose to model the specific situation of heat transfer via surface polaritons between two thick dielectric slabs separated by a planar vacuum gap of thickness~$d$, so that the half-spaces $z>d/2$ and $z<-d/2$ are occupied by dielectrics with a dielectric function $\varepsilon(\omega)$ and held at temperatures $T_1$ and $T_2$, respectively. The circuit mimicking this situation was studied in detail in Sec.~\ref{sec:circuit_polaritons}.
For a metallic system, the Green's functions' treatment is analogous, and can be found in Ref.~\cite{Wise:2022}.

\subsection{Microscopic model}

We need a microscopic model for a dielectric that would produce an isotropic Drude-Lorentz dielectric function of the bulk material,
\begin{equation}\label{eq:Drude-Lorentz}
\varepsilon(\omega) = \varepsilon_\infty\,\frac{\omega_L^2-\omega^2-2i\gamma\omega}{\omega_T^2-\omega^2-2i\gamma\omega},
\end{equation}
where the spatial dispersion is neglected.
Since our goal is to illustrate the Green's function approach rather than to quantitatively describe a specific material, we aim at the simplest model with a minimal number of ingredients. When a specific example is needed, we will have in mind the tetrahedral SiC with two atoms per unit cell (3C-SiC) and point group~$T_d$. For simplicity, we will set the high-frequency value of the dielectric function $\varepsilon_\infty=1$. Its difference from unity is due to a contribution from high-frequency polarizable degrees of freedom. We do not include them explicitly, but briefly discuss the procedure in the end of Sec.~\ref{sec:phonon_propagators}.

\begin{subequations}\label{eqs:Hamiltonian}

The first ingredient of the model is the triply degenerate polar optical phonon mode, assumed to be dispersionless. It is described by a vector field $\boldsymbol\xi(\vec{r})$ of the relative displacements of two inequivalent atoms, with the Hamiltonian
\begin{equation}\label{eq:Hop=}
\hat{H}_\text{op} = \int_{|z|>d/2}{d}^3\vec{r}\left[\frac{\hat\eta_j(\vec{r})\,\hat\eta_j(\vec{r})}{2\varrho_\text{op}} + \frac{\varrho_\text{op}\omega_T^2}2\,\hat\xi_j(\vec{r})\,\hat\xi_j(\vec{r})\right].
\end{equation}
Here $\hat{\boldsymbol\eta}(\vec{r})$ is the momentum canonically conjugate to the displacement operator $\hat{\boldsymbol\xi}(\vec{r})$, so that $[\hat\xi_j(\vec{r}),\hat\eta_k(\vec{r}')]=i\hbar\delta_{jk}\delta(\vec{r}-\vec{r}')$. Summation over repeated Cartesian indices $j,k=x,y,z$ is implied. $\varrho_\text{op}$~is the density of the reduced mass, corresponding to the optical phonon. For 3C-SiC, it is related to the total mass density $\varrho_0$ of the crystal by a factor involving the masses of silicon and carbon atoms, $\varrho_\text{op}=\varrho_0m_\text{Si}m_\text{C}/(m_\text{Si}+m_\text{C})^2$. Finally, $\omega_T$ is the mechanical frequency of the optical phonon.

The second ingredient is the coupling of optical phonons to the electromagnetic field. It is due to the fact that the polar phonon displacement $\boldsymbol\xi(\vec{r})$ is associated with electric polarization in the material, $\vec{P}(\vec{r})=\theta(|z|-d/2)\,\rho\,\boldsymbol\xi(\vec{r})$, with $\theta(z)$~being the Heaviside step function. Such local relation is valid under the same assumptions as neglecting the spatial dispersion in the dielectric function~(\ref{eq:Drude-Lorentz}), that is, at length scales well exceeding the atomic scale. The coefficient $\rho$ has the dimensionality of the charge density and is determined by the microscopic charge distribution in the atomic bond. Coupling between the polarization and the electromagnetic field is taken in the electrostatic (Coulomb) limit, valid for heat transfer between dielectrics at distances much smaller than the wavelength of thermal photons (in contrast to metals, where the dominant mechanism is magnetostatic~\cite{Polder:1971,Chapuis:2008.1,Chapuis:2008.2,Wise:2021near}). Recalling that the charge density is given by $-\grad\cdot\hat{\vec{P}}(\vec{r})$, we write
\begin{align}
    \hat{H}_\text{C} = {}&{}  \frac{1}2\int d^3\vec{r}\,d^3\vec{r}'\,\frac{(-\grad\cdot\hat{\vec{P}}(\vec{r}))\,(-\grad'\cdot\hat{\vec{P}}(\vec{r}'))}{|\vec{r}-\vec{r}'|} \nonumber\\
\equiv {}&{} \frac{\rho^2}2\int d^3\vec{r}\,d^3\vec{r}'\,D_{ij}(\vec{r}-\vec{r}')\,\hat{P}_i(\vec{r})\,\hat{P}_j(\vec{r}').
\label{eq:HC=}
\end{align}
With this definition, the function $-D_{ij}(\vec{r}-\vec{r}')$ gives the response of the electric field to the polarization, as determined by Maxwell's equations in the quasi-static limit, and describes the dipole-dipole interaction.

The third ingredient has to account for the optical phonon damping~$\gamma$, appearing as the imaginary term in the dielectric function~(\ref{eq:Drude-Lorentz}). Such damping can result from optical phonon decay into a pair of acoustic phonons due to anharmonicity~\cite{Cowley:1965, Klemens:1966}, provided that the optical phonon frequency is not too high and can be matched by two acoustic phonon frequencies (which is the case in SiC~\cite{Kushwaha:1982, Lee:1982, Wang:2017}). The acoustic phonons are described by the vector field of atomic displacements $\vec{u}(\vec{r})$. In the long-wavelength limit, the change in the elastic energy is expressed in terms of the strain tensor $u_{ij}\equiv(1/2)(\partial{u}_i/\partial{x}_j+\partial{u}_j/\partial{x}_i)$~\cite{Landafshitz:VII}. The simplest model for acoustic phonons is that of a fully isotropic elastic medium; however, such high symmetry prohibits local anharmonic coupling between a vector~$\xi_i$ (transforming according to angular momentum~1) and a product of two symmetric tensors~$u_{ij}$ (each transforming according to angular momentum 0 or 2, so their product having components of angular momentum 0, 2, and~4). Thus, we write the elastic energy in terms of invariants of the $T_d$~point group, whose lower symmetry allows for the sought anharmonic coupling in the form
\begin{align}
    \hat{H}_\text{anh} = \Upsilon\int_{|z|>d/2}{d}^3\vec{r}
\left[\hat\xi_x\hat{u}_{yz}(2\hat{u}_{xx}-\hat{u}_{yy}-\hat{u}_{zz})\right.
\nonumber\\
{}+\hat\xi_y\hat{u}_{zx}(2\hat{u}_{yy}-\hat{u}_{zz}-\hat{u}_{xx})\nonumber\\
{}+\left.\hat\xi_z\hat{u}_{xy}(2\hat{u}_{zz}-\hat{u}_{xx}-\hat{u}_{yy})\right]\nonumber\\
\equiv\int_{|z|>d/2}{d}^3\vec{r}\,\frac{\Upsilon_{ijklm}}2\,\hat\xi_i(\vec{r})\,\hat{u}_{jk}(\vec{r})\,\hat{u}_{lm}(\vec{r}),
\label{eq:Hanh=}
\end{align}
where $\Upsilon$ is the corresponding coupling constant, and the $x,y,z$ axes are directed along the edges of the cube centered at the position of a Si atom, while the four C atoms forming the tetrahedron are located at four (out of eight) cube's vertices, so that the six edges of the tetrahedron coincide with diagonals of each of the six cube's faces.

The harmonic part of the acoustic phonon Hamiltonian is written in terms of the strain tensor operator $\hat{u}_{ij}(\vec{r})$ and of the vector momentum operator $\hat{\boldsymbol\pi}(\vec{r})$, canonically conjugate to the displacement operator $\hat{\vec{u}}(\vec{r})$, such that $[\hat{u}_j(\vec{r}),\hat\pi_k(\vec{r}')]=i\hbar\delta_{jk}\delta(\vec{r}-\vec{r}')$. To be consistent, here we also assume the tetrahedral symmetry instead of the full isotropy. This gives three terms in the elastic deformation energy:
\begin{align}
\hat{H}_\text{ac} = {}&{}\int_{|z|>d/2}{d}^3\vec{r}\left[\frac{\hat\pi_i(\vec{r})\,\hat\pi_i(\vec{r})}{2\varrho_0} 
+ \varrho_0\,\frac{v_l^2-2v_t^2}2\left[\hat{u}_{ii}(\vec{r})\right]^2\right.\nonumber\\
{}&{}  \hspace*{1.7cm} + \varrho_0v_t^2\hat{u}_{ij}(\vec{r})\hat{u}_{ij}(\vec{r}) \nonumber\\
{}&{} \hspace*{1.7cm} +\left. \frac{\varrho_0v_\Delta^2}2\left[\hat{u}_{xx}^2(\vec{r})+\hat{u}_{yy}^2(\vec{r})+\hat{u}_{zz}^2(\vec{r})\right]\right]\nonumber\\
\equiv{}&{}\int_{|z|>d/2}{d}^3\vec{r}\left[\frac{\hat\pi_i(\vec{r})\,\hat\pi_i(\vec{r})}{2\varrho_0}+\frac{\mathcal{K}_{jklm}}2\,\hat{u}_{jk}(\vec{r})\,\hat{u}_{lm}(\vec{r})\right].
\label{eq:Hac=}
\end{align}
Here $\varrho_0$ is the mass density of the crystal, $v_l$~and~$v_t$ are the longitudinal and transverse acoustic phonon velocities in the isotropic approximation (which corresponds to keeping only the first two terms in the elastic energy), while $v_\Delta$ plays the role of the coupling constant in the last term, which is absent in the isotropic approximation. This last term splits the two transverse phonon branches for wave vectors which are not directed along high-symmetry axes.

\end{subequations}

\subsection{Phonon propagators and the dielectric function}
\label{sec:phonon_propagators}

In order to verify that the model of Eqs.~(\ref{eqs:Hamiltonian}) indeed corresponds to the dielectric function~(\ref{eq:Drude-Lorentz}) with $\varepsilon_\infty=1$, let us consider an infinite bulk crystal and calculate the response of the polarization to an external oscillating electric field $\vec{E}^\text{ext}e^{i\vec{k}\vec{r}-i\omega{t}}+\mbox{c.c.}$, which couples to the polarization via the Hamiltonian
\begin{equation}
    \hat{H}_\text{ext} = -\int{d}^3\vec{r}\,\hat{\vec{P}}(\vec{r})\cdot\left(\vec{E}^\text{ext}e^{i\vec{k}\vec{r}-i\omega{t}}+\mbox{c.c.}\right).
\end{equation}
The Kubo formula expresses the corresponding response function $\tilde\chi_{ij}(\vec{k},\omega)$ as the retarded correlator of the polarizations:
\begin{equation}\label{eq:Kubo}
    \tilde\chi_{ij}(\vec{k},\omega) = \frac{i}\hbar\int\limits_0^\infty{d}t\int{d}^3\vec{r}\,e^{i\omega{t}-i\vec{k}\vec{r}}\left\langle[\hat{P}_i(\vec{r},t),\hat{P}_j(\boldsymbol{0},0)]\right\rangle,
\end{equation}
where we used the fact that in the infinite medium two-point correlators depend only on the difference of the spatial and temporal coordinates.

To evaluate this correlator, we treat $\hat{H}_\text{op}+\hat{H}_\text{ac}$ [Eqs.~(\ref{eq:Hop=}) and~(\ref{eq:Hac=})] as the zero-approximation Hamiltonian, while $\hat{H}_\text{C}$ and $\hat{H}_\text{anh}$ [Eqs.~(\ref{eq:HC=}) and~(\ref{eq:Hanh=}), respectively] are included by summing perturbation series. Accordingly, the operators of the optical and acoustic phonon displacements are expressed in terms of the phonon creation and annihilation operators:
\begin{subequations}\begin{align}
    &\hat\xi_j(\vec{r}) = \sum_{\vec{k}}\sqrt{\frac{\hbar}{2V\varrho_\text{op}\omega_T}}\left(\hat{b}_{\vec{k}j}e^{i\vec{k}\vec{r}} + \hat{b}^\dagger_{\vec{k}j}e^{-i\vec{k}\vec{r}}\right),\\
    &\hat{\vec{u}}(\vec{r}) = \sum_{\vec{k},\mu}\sqrt{\frac{\hbar}{2V\varrho_0\omega^\text{ac}_{\vec{k},\mu}}}\,\vec{e}_{\vec{k}\mu}\left(\hat{a}_{\vec{k}\mu}e^{i\vec{k}\vec{r}} + \hat{a}^\dagger_{\vec{k}\mu}e^{-i\vec{k}\vec{r}}\right),
\end{align}\end{subequations}
where $V$ is the crystal volume. The index $\mu$ labels the three acoustic phonon branches, while the degenerate optical phonon branches can be labeled by the Cartesian index $j=x,y,z$. The unit vectors $\vec{e}_{\vec{k}\mu}$ give the directions of the atomic displacements for each acoustic normal mode. The corresponding non-interacting retarded, advanced, and Keldysh Green's functions (propagators) for the optical phonons are given by
\begin{subequations}\begin{align}
G_{jl}^R(\vec{r}-\vec{r}',t-t') = {}&{} -\frac{i}\hbar\,\theta(t-t') \langle[\hat\xi_j(\vec{r},t),\hat\xi_l(\vec{r}',t')]\rangle
\nonumber\\
={}&{}\int\frac{d^3\vec{k}\,d\omega}{(2\pi)^4}\,
\frac{\delta_{jl}}{\varrho_\text{op}}
\frac{e^{i\vec{k}(\vec{r}-\vec{r}')-i\omega(t-t')}}{(\omega+i0^+)^2-\omega_T^2},\\
G_{jl}^A(\vec{r}-\vec{r}',t-t')  = {}&{}\frac{i}\hbar\,\theta(t'-t) \langle[\hat\xi_j(\vec{r},t),\hat\xi_l(\vec{r}',t')]\rangle
\nonumber\\
{}&{}=\int\frac{d^3\vec{k}\,d\omega}{(2\pi)^4}\,
\frac{\delta_{jl}}{\varrho_\text{op}}
\frac{e^{i\vec{k}(\vec{r}-\vec{r}')-i\omega(t-t')}}{(\omega-i0^+)^2-\omega_T^2},\\
G_{jl}^K(\vec{r}-\vec{r}',t-t') = {}&{} -\frac{i}\hbar\langle\{\hat\xi_j(\vec{r},t),\hat\xi_l(\vec{r}',t')\}\rangle
\nonumber\\
={}&{}\int\frac{d^3\vec{k}\,d\omega}{(2\pi)^4}\,
\frac{\delta_{jl}}{\varrho_\text{op}}
e^{i\vec{k}(\vec{r}-\vec{r}')-i\omega(t-t')}\frac{2\pi{i}}{2\omega_T}\nonumber\\
{}&{}\times[\delta(\omega+\omega_T)-\delta(\omega-\omega_T)]\coth\frac{\hbar\omega}{2T},
\end{align}\end{subequations}
where $\{\ldots\}$ denotes the anticommutator, and $T$ is the crystal temperature.
In the similar way we define the Green's functions $U^{R,A,K}_{jklm}(\vec{r}-\vec{r}',t-t')$ for the acoustic phonons, building them from the strain operators $\hat{u}_{jk}(\vec{r},t),\hat{u}_{lm}(\vec{r}',t')$. These Green's functions can be straightforwardly evaluated in the Fourier space from the Hamiltonian~(\ref{eq:Hac=}); we do not give the explicit expressions because they are too cumbersome.

\begin{figure}
    \centering
    \includegraphics[width=0.95\linewidth]{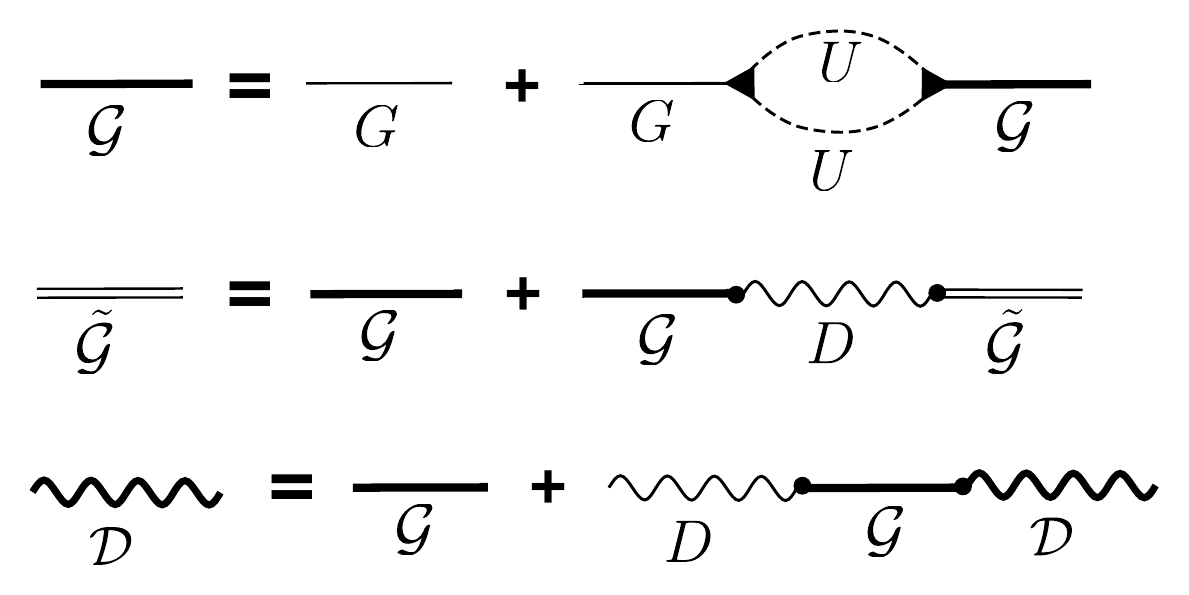}
    \caption{Feynamnn diagrams representing the Dyson series for the phonon Green's functions. Solid, dashed and wavy lines represent the Green's function for the optical phonons, acoustic phonons, and the electric field, respectively. Triangles and thick dots represent the $\Upsilon$ and $\rho$ vertices appearing in the Hamiltonians $\hat{H}_\text{anh}$ and $\hat{H}_\text{C}$, respectively.}
    \label{fig:diag_propagators}
\end{figure}

The next step is to dress the optical phonon Green's function by the self-energy originating from the coupling~(\ref{eq:Hanh=}) to the acoustic phonons (Fig.~\ref{fig:diag_propagators}). To the leading (i.~e, second) order in $\hat{H}_\text{anh}$, the retarded self-energy is given by
\begin{align}
    \Sigma_{jj'}^R(\vec{K},\Omega) = {}&{}
    \int\frac{d^3\vec{k}\,d\omega}{(2\pi)^4}\,\frac{i\hbar}{4}\,
    \Upsilon_{jklmn}\Upsilon_{j'k'l'm'n'}\nonumber\\
    {}&{}\times\left[
    U_{klk'l'}^R(\vec{K}+\vec{k},\Omega+\omega)\,U_{m'n'mn}^K(\vec{k},\omega)\right.\nonumber\\
    {}&{}+\left.U_{klk'l'}^K(\vec{K}+\vec{k},\Omega+\omega)\,U_{m'n'mn}^A(\vec{k},\omega)\right].
\end{align}
This self-energy can be calculated explicitly using the Green's functions built from the Hamiltonian~(\ref{eq:Hac=}). Note, however, that what we really need is the limit $\Im\Sigma_{jj'}^R(\vec{k}\to0,\omega)$, because we are interested in the damping of long-wavelength optical phonons. By symmetry, $\Sigma_{jj'}^R(\vec{k}\to0,\omega)\propto\delta_{jj'}$, so its real part gives an overall frequency shift which is the same for the three phonon branches, without lifting the degeneracy, so it can be absorbed into~$\omega_T$. Thus, we approximate
\begin{equation}
    \Sigma_{jl}^R(\vec{k},\omega)=-2i\varrho_\text{op}\omega\,\gamma(\omega)\,\delta_{jl},
\end{equation}
where the damping rate $\gamma(\omega)$ is frequency-dependent. This frequency dependence is determined by the two-phonon density of states for acoustic phonons, as well as by the momentum dependence of the coupling matrix elements. Equations~(\ref{eq:Hac=}) and~(\ref{eq:Hanh=}), although provide a specific model where $\gamma(\omega)$ can be calculated, do not describe any material quantitatively. The reason is that an optical phonon with small momentum decays into two acoustic phonons whose momenta are not small at all, they are of the order of the inverse lattice constant. Equations~(\ref{eq:Hac=}) and~(\ref{eq:Hanh=}) represent an expansion in the acoustic phonon momenta around $\vec{k}=0$, and lose their validity deep in the Brillouin zone. Thus, we keep $\gamma(\omega)$ as a parameter of the theory (the simplest assumption, implied in the Drude-Lorentz model, is to take $\gamma=\text{const}$ in the frequency region of interest). Summing up the Dyson series, we find the optical phonon Green's function, dressed by the anharmonic coupling to acoustic phonons:
\begin{equation}\label{eq:mathcalG=}
    \mathcal{G}^R_{jl}(\vec{k},\omega)=\frac{\delta_{jl}}{\varrho_\text{op}}\, \frac{1}{\omega^2-\omega_T^2+2i\omega\gamma(\omega)}.
\end{equation}

Finally, we dress this Green's function by the self-energy originating from the Coulomb interaction~(\ref{eq:HC=}), taken for the infinite medium:
\begin{equation}
    \tilde\Sigma_{jl}(\vec{k}) = \rho^2D_{jl}(\vec{k}) = 4\pi\rho^2\,\frac{k_jk_l}{k^2}.
\end{equation}
This gives the fully dressed optical phonon retarded Green's function  (Fig.~\ref{fig:diag_propagators}),
\begin{align}
    \tilde{\mathcal{G}}^R_{jl}(\vec{k},\omega) = {}&{}
    \left(\delta_{jl}-\frac{k_jk_l}{k^2}\right)\,\frac{1/\varrho_\text{opt}}{\omega^2-\omega_T^2+2i\omega\gamma(\omega)} \nonumber\\
    {}&{} + \frac{k_jk_l}{k^2}\,\frac{1/\varrho_\text{opt}}{\omega^2-\omega_L^2+2i\omega\gamma(\omega)},
\end{align}
where $\omega_L\equiv\sqrt{\omega_T^2+4\pi\rho^2/\varrho_\text{op}}$ is the Coulomb frequency of the longitudinal optical phonon.

To relate the Green's functions to the dielectric function~$\varepsilon(\omega)$, we recall that the latter determines the response of the electric displacement $\vec{E}+4\pi\vec{P}$ to the \emph{total} electric field $\vec{E}$, which is the sum of the external field~$\vec{E}^\text{ext}$ and the field produced by the polarization~$\vec{P}$ itself, so that $E_j=E_j^\text{ext}-D_{jl}P_l = (\delta_{jm}-D_{jl}\tilde\chi_{lm})E_m^\text{ext}$. We can also define the electric susceptibility $\chi(\vec{k},\omega)$, which determines the response of the polarization $\vec{P}$ to the total field~$\vec{E}$, so that the dielectric function $\varepsilon_{ij}=\delta_{ij} + 4\pi\chi_{ij}$. Equating $\tilde\chi\vec{E}^\text{ext} = \chi(\vec{E}^\text{ext}-D\tilde\chi\vec{E}^\text{ext})$, we find the relation $\chi=\tilde\chi(1-D\tilde\chi)^{-1}$. At the same time, Eq.~(\ref{eq:Kubo}) gives $\tilde\chi_{ij}(\vec{k},\omega)=-\rho^2\tilde{\mathcal{G}}^R_{ij}(\vec{k},\omega)$, and the Dyson series for $\tilde{\mathcal{G}}^R$ yields $(\tilde{\mathcal{G}}^R)^{-1} = ({\mathcal{G}}^R)^{-1}-\rho^2D$, from which we deduce $\chi_{ij}(\vec{k},\omega)=-\rho^2{\mathcal{G}}^R_{ij}(\vec{k},\omega)$. As a result, from Eq.~(\ref{eq:mathcalG=}) with $\gamma(\omega)=\text{const}$ we recover Eq.~(\ref{eq:Drude-Lorentz}) with $\varepsilon_\infty=1$. Thus, the Hamiltonian presented in the previous subsection indeed describes a dielectric material with the Drude-Lorentz dielectric function.

To reproduce Eq.~(\ref{eq:Drude-Lorentz}) with an asymptotic high-frequency value~$\varepsilon_\infty>1$, one has to include additional polarizable degrees of freedom resonating at much higher frequencies (for example, electronic contribution to atomic polarizabilities). The propagator for these degrees of freedom should be included (i)~as an additive contribution to $\tilde\chi_{ij}(\omega)$, and (ii)~in the dressing of the dipole-dipole interaction $D_{ij}(\vec{r}-\vec{r}')$. Then the Drude-Lorentz dielectric function with a single resonance at $\omega=\omega_T$ is recovered for frequencies well below those of the higher resonances.

\subsection{Energy current operator}

To define the energy current operator $\hat{\vec{J}}(\vec{r})$, one must first define the energy density operator $\hat{\mathcal{E}}(\vec{r})$, such that the Hamiltonian $\hat{H}=\int{d}^3\vec{r}\,\hat{\mathcal{E}}(\vec{r})$, and then define an operator $\hat{\vec{J}}(\vec{r})$ which satisfies the energy continuity equation,
\begin{equation}\label{eq:energy_continuity}
\frac{\partial\hat{\mathcal{E}}(\vec{r})}{\partial{t}}=\frac{i}\hbar\left[\hat{H},\hat{\mathcal{E}}(\vec{r})\right]=-\grad\cdot\hat{\vec{J}}(\vec{r}).
\end{equation}
The energy densities $\hat{\mathcal{E}}_\text{op}(\vec{r}), \hat{\mathcal{E}}_\text{ac}(\vec{r}), \hat{\mathcal{E}}_\text{anh}(\vec{r})$ are naturally defined as the integrands in Eqs.~(\ref{eq:Hop=}), (\ref{eq:Hac=}) and (\ref{eq:Hanh=}), respectively.
Since the Coulomb energy~(\ref{eq:HC=}) is non-local, it is possible to find many operators $\hat{\mathcal{E}}_\text{C}(\vec{r})$ which give $\hat{H}_\text{C}=\int{d}^3\vec{r}\,\hat{\mathcal{E}}_\text{C}(\vec{r})$. The gauge-invariant definition is~\cite{Catelani:2005}
\begin{align}
    \hat{\mathcal{E}}(\vec{r}) = {}&{} \left[\hat{\mathcal{E}}_\text{op}(\vec{r}) + \hat{\mathcal{E}}_\text{ac}(\vec{r}) + \hat{\mathcal{E}}_\text{anh}(\vec{r})\right]\theta(|z|-d/2) \nonumber\\ 
    {}&{} + \frac{1}{8\pi}\left|\int{d}^3\vec{r}'\,\grad\frac{1}{|\vec{r}-\vec{r}'|}\grad'\cdot\rho\hat{\boldsymbol\xi}(\vec{r}')\theta(|z'|-d/2)\right|^2.
    \label{eq:calE=}
\end{align}
Indeed, the identity
\begin{equation}
    -\nabla^2\frac{1}{|\vec{r}-\vec{r}'|} = 4\pi\delta(\vec{r}-\vec{r}')
\end{equation}
ensures that the integral over $\vec{r}$ of the last term in Eq.~(\ref{eq:calE=}) gives Eq.~(\ref{eq:HC=}).
Then, the energy current operator $\hat{\vec{J}}(\vec{r})$ satisfying Eq.~(\ref{eq:energy_continuity}) is given by
\begin{align}\label{eq:J=}
    &\hat{J}_i(\vec{r}) = \frac{1}2\left\{\hat\varsigma_{ik}(\vec{r}),\frac{\hat\pi_k(\vec{r})}{\varrho_0}\right\} \theta(|z|-d/2)\nonumber\\
    &\hspace*{1cm} {} + \frac{\hat{\vec{E}}(\vec{r})\times\hat{\boldsymbol{\beta}}(\vec{r})-\hat{\boldsymbol{\beta}}(\vec{r})\times\hat{\vec{E}}(\vec{r})}{8\pi}\equiv\hat{J}^\varsigma_i+\hat{J}^S_i,
\end{align}
where the first line represents the elastic energy current, expressed in terms of the stress tensor,
\begin{equation}
\hat\varsigma_{ik}(\vec{r})=\left(\Lambda_{iklm}+\hat\xi_j(\vec{r})\Upsilon_{jiklm}\right)\hat{u}_{lm}(\vec{r}),
\end{equation}
while the second line in Eq.~(\ref{eq:J=}) is nothing but the Poynting vector, expressed in terms of the electric and ``magnetic'' fields,
\begin{subequations}\label{eqs:E=b=}\begin{align}
    &\hat{\vec{E}}(\vec{r})=-\grad\int{}d^3\vec{r}'\,\frac{-\grad'\cdot\rho\hat{\boldsymbol\xi}(\vec{r}')\theta(|z'|-d/2)}{|\vec{r}-\vec{r}'|},\\
    &\hat{\boldsymbol{\beta}}(\vec{r})=\grad\times\int{}d^3\vec{r}'\,\frac{(\rho/\varrho_\text{op})\hat{\boldsymbol\eta}(\vec{r}')\theta(|z'|-d/2)}{|\vec{r}-\vec{r}'|},
\end{align}\end{subequations}
which satisfy the following equations:
\begin{subequations}\label{eqs:Maxwell}
\begin{align}
&\grad\cdot\hat{\vec{E}}(\vec{r})=-4\pi\grad\cdot\hat{\vec{P}}(\vec{r}),\\
&\grad\times\hat{\vec{E}}(\vec{r})=0,\label{eq:Faraday}\\
&\grad\cdot\hat{\boldsymbol{\beta}}(\vec{r})=0,\\
&\grad\times\hat{\boldsymbol{\beta}}(\vec{r})=4\pi\,\frac{i}\hbar\left[\hat{H},\hat{\vec{P}}(\vec{r})\right]+\frac{i}\hbar\left[\hat{H},\hat{\vec{E}}(\vec{r})\right].
\end{align}\end{subequations}
The first two equations are the Maxwell's equations for the electrostatic electric field $\vec{E}$, while the last two equations are the Maxwell's equations for the magnetic field~$\vec{B}$, up to the factor $1/c$ on the right-hand side of the last equation. Indeed, the two terms with the commutators in the last equation become $4\pi(\partial\hat{\vec{P}}/\partial{t})\equiv4\pi\vec{j}$ and $\partial\hat{\vec{E}}/\partial{t}$ in the Heisenberg representation, where 
\begin{align}
    \hat{\vec{j}}(\vec{r}){}&{} =
    \frac\rho{\varrho_\text{op}}\,\hat{\boldsymbol\eta}(\vec{r}) \nonumber\\
    {}&{} =
    -i\omega_T\sum_{\vec{k},l}\sqrt{\frac{\hbar\rho^2}{2V\varrho_\text{op}\omega_T}}\,\vec{e}_l\left(\hat{b}_{\vec{k}l}e^{i\vec{k}\vec{r}} - \hat{b}^\dagger_{\vec{k}l}e^{-i\vec{k}\vec{r}}\right)
\end{align}
is the operator of the electric current density.

The appearance of the magnetic field may look somewhat surprising in our problem with purely electrostatic interactions. Strictly speaking, the electrostatic limit is obtained by sending the speed of light $c\to\infty$ in the Maxwell's equations, so the magnetic field $\vec{B}=O(1/c)$ vanishes, but the ``magnetic'' filed $\boldsymbol{\beta}=c\vec{B}$ stays finite, and so does the Poynting vector. The only term in the Maxwell's equations that vanishes in the limit $c\to\infty$ is $-(1/c^2)\partial\hat{\boldsymbol{\beta}}/\partial{t}$, which would stand on the right-hand side of Eq.~(\ref{eq:Faraday}).

Using Eqs.~(\ref{eqs:Maxwell}), it is rather straightforward to verify that the continuity equation~(\ref{eq:energy_continuity}) is satisfied. Indeed, we have
\begin{align*}
&    [\hat{H}_\text{op}+\hat{H}_\text{anh},\hat{\mathcal{E}}_\text{op}+\hat{\mathcal{E}}_\text{anh}] = 0,\\
& [\hat{H}_\text{C},\hat{\mathcal{E}}_\text{anh}] = [\hat{H}_\text{anh}+\hat{H}_\text{C},\hat{E}^2/(8\pi)] = 0,\\
& \frac{i}\hbar[\hat{H}_\text{ac},\hat{\mathcal{E}}] + \frac{i}\hbar[\hat{H}_\text{op}+\hat{H}_\text{anh}+\hat{H}_\text{C},\hat{\mathcal{E}}_\text{ac}]= -\grad\cdot\hat{\vec{J}}^\varsigma,\\
&\frac{i}\hbar[\hat{H}_\text{op},\hat{E}^2/(8\pi)]=-\grad\cdot\hat{\vec{J}}^S-\frac{\hat{\vec{j}}\cdot\hat{\vec{E}}+\hat{\vec{E}}\cdot\hat{\vec{j}}}2,\\
&\frac{i}\hbar[\hat{H}_\text{C},\hat{\mathcal{E}}_\text{op}]=\frac{\hat{\vec{j}}\cdot\hat{\vec{E}}+\hat{\vec{E}}\cdot\hat{\vec{j}}}2.
\end{align*}

\subsection{Polarization operator}
\label{App:PolarizationOp}

Instead of working with the phonon Green's functions, it is convenient to introduce retarded, advanced and Keldysh components of the polarization operator,
\begin{subequations}\label{eqs:tildePi=}\begin{align}
\tilde\Pi_{jk}^R(\vec{r},\vec{r}',t-t') = {}&{} -\frac{i}\hbar\,\theta(t-t') \langle[\hat{P}_j(\vec{r},t),\hat{P}_k(\vec{r}',t')]\rangle,\\
\tilde\Pi_{jk}^A(\vec{r},\vec{r}',t-t')  = {}&{}\frac{i}\hbar\,\theta(t'-t) \langle[\hat{P}_j(\vec{r},t),\hat{P}_k(\vec{r}',t')]\rangle,\\
\tilde\Pi_{jk}^K(\vec{r},\vec{r}',t-t') = {}&{} -\frac{i}\hbar\langle\{\hat{P}_j(\vec{r},t),\hat{P}_k(\vec{r}',t')\}\rangle,
\end{align}\end{subequations}
where the time dependence is determined by the full Hamiltonian.
These are proportional to the corresponding Green's functions of the displacements  $\hat{\boldsymbol\xi}(\vec{r})$: $\tilde\Pi_{jk}^{R,A,K}(\vec{r},\vec{r}',t-t')=\rho^2\tilde{\mathcal{G}}_{jk}^{R,A,K}(\vec{r},\vec{r}',t-t')$. The Green's functions of the momenta $\hat{\boldsymbol\eta}(\vec{r})$ can also be expressed in terms of the polarization operator using the fact that $\hat{\boldsymbol\eta}(\vec{r},t)=\varrho_\text{op}\,\partial\hat{\boldsymbol\xi}(\vec{r},t)/\partial{t}$:
\begin{subequations}\begin{align}
{}&{} \mp\frac{i}\hbar\,\theta(\pm(t-t')) \langle[\hat{\eta}_j(\vec{r},t),\hat{\eta}_k(\vec{r}',t')]\rangle \nonumber\\
{}&{}= \frac{\varrho_\text{op}^2}{\rho^2}{}\frac{\partial^2\tilde\Pi_{jk}^{R,A}(\vec{r},\vec{r}',t-t')}{\partial{t}\,\partial{t}'} \nonumber \\
{}&{} \quad {} - \rho_\text{op}\delta_{jk}\delta(t-t')\delta(\vec{r}-\vec{r}')\theta(|z|-d/2),\\
{}&{} -\frac{i}\hbar\langle\{\hat{\eta}_j(\vec{r},t),\hat{\eta}_k(\vec{r}',t')\}\rangle= \frac{\varrho_\text{op}^2}{\rho^2}{}\frac{\partial^2\tilde\Pi_{jk}^K(\vec{r},\vec{r}',t-t')}{\partial{t}\,\partial{t}'}.
\end{align}\end{subequations}
The reason for introducing the polarization operator rather than working directly with the phonon Green's functions is that the subsequent steps will not rely on the specific model for the material that we used in the previous subsections. Whatever excitations contribute to the dielectric response of the material, the subsequent steps involve only the polarization operator, and thus are more general than the specific model. For example, for a metal the polarization operator is expressed in the random-phase approximation in terms of density-density and current-current electronic bubbles, and the whole subsequent treatment is quite analogous~\cite{Wise:2022}.

Together with the full polarization operator $\tilde\Pi$ built from operators in the Heisenberg representation involving the full Hamiltonian including interaction~(\ref{eq:HC=}) with the electric field, we also introduce the mechanical polarization operator~$\Pi$, where the time dependence does not include this interaction. It is related to the optical phonon Green's functions as $\Pi_{jk}^{R,A,K}(\vec{r},\vec{r}',t-t')=\rho^2{\mathcal{G}}_{jk}^{R,A,K}(\vec{r},\vec{r}',t-t')$, and in the frequency space is given by
\begin{subequations}\label{eqs:PiKRA=}\begin{align}
\Pi_{jk}^R(\vec{r},\vec{r}',\omega) = {}&{} \delta_{jk}\delta(\vec{r}-\vec{r}')\,\frac{1-\varepsilon(\omega)}{4\pi}\,\theta(|z|-d/2),
\label{eq:PiR=}\\
\Pi_{jk}^A(\vec{r},\vec{r}',\omega) = {}&{} \delta_{jk}\delta(\vec{r}-\vec{r}')\,\frac{1-\varepsilon^*(\omega)}{4\pi}\,\theta(|z|-d/2),\\
\Pi_{jk}^K(\vec{r},\vec{r}',\omega) = {}&{} \delta_{jk}\delta(\vec{r}-\vec{r}')\,\frac{\varepsilon^*(\omega)-\varepsilon(\omega)}{4\pi}\nonumber\\
{}&{}\times \left[\theta(z-d/2)\coth\frac{\hbar\omega}{2T_1} \right.
\nonumber\\ {}&{}\quad 
+ \left.\theta(-z-d/2)\coth\frac{\hbar\omega}{2T_2}\right].
\end{align}\end{subequations}
These expressions are also more general than the specific model for the material that we used in the previous subsections~\cite{Abrikosov:1975}. The conditions for their validity coincide with those for the spatially local dielectric response of the material, supplemented by the requirement of thermal equilibrium in the material on each side of the vacuum gap. In other words, heat conduction inside the material should be more efficient than the near-field radiative heat transfer across the vacuum gap~\cite{Reina:2020,Reina:2021}. The relation between the mechanical and the Coulomb-dressed polarization operators at a given frequency~$\omega$ is
\begin{subequations}\begin{align}
&\tilde\Pi^{R,A} = \Pi^{R,A} + \Pi^{R,A}\, D\, \tilde\Pi^{R,A},\\
&\tilde\Pi^R-\tilde\Pi^A = \left(\openone + \Pi^R\mathcal{D}^R\right)\left(\Pi^R-\Pi^A\right)\left(\openone+\mathcal{D}^A\Pi^A\right),\label{eq:tildePiRA=}\\
&\tilde\Pi^K = \left(\openone + \Pi^R\mathcal{D}^R\right)\Pi^K\!\left(\openone+\mathcal{D}^A\Pi^A\right),\label{eq:tildePiK=}
\end{align}
where the products are understood in the matrix sense (convolutions over the Cartesian indices and over the spatial variables), $\openone$ is a shorthand notation for $\delta_{ij}\delta(\vec{r}-\vec{r}')$, and the dressed retarded and advanced Green's functions of the electric field are defined as (Fig.~\ref{fig:diag_propagators})
\begin{equation}
    \mathcal{D}^{R,A} = D + D\,\Pi^{R,A}\mathcal{D}^{R,A}.
\end{equation}
\end{subequations}
While $-D$ gives the electric field produced by an external polarization, as determined by Maxwell's equations in the empty space, the dressed Green's function $-\mathcal{D}^R$ does the same, but from the solution of Maxwell's equations in the full dielectric environment. This dielectric environment is encoded in~$\Pi^R$~\cite{Abrikosov:1975,Landafshitz:IX}.

\subsection{Heat-current fluctuations}

As in most of the literature on near-field radiative heat transfer, we will be interested in the energy current $\hat{J}_z(\vec{r})$ inside the vacuum gap separating the two materials, where the Poynting vector represents the only contribution. Specifically, Ref.~\cite{Biehs:2018} focused on fluctuations of $\hat{J}_z(\vec{r})$ at a given point $\vec{r}=(0,0,d/2-0^+)$. We believe that a more relevant quantity would be the total power flowing to the upper body, assumed to have a finite size $L_x\times{L}_y$ in the in-plane directions:
\begin{equation}
    \hat{\mathcal{P}}_1=\int_0^{L_x}dx\int_0^{L_y}dy\,\hat{J}_z(x,y,d/2-0^+).
\end{equation}
Indeed, in experiments it is not the heat current that is measured directly, but rather fluctuations in the sample's temperature~\cite{Karimi:2020}. Conversion of the heat current into a temperature change necessarily involves heat transport inside the material, which is assumed to be efficient. Thus, local fluctuations of the heat current will quickly spread out across the sample, so fluctuations of the global energy stored in the finite-size sample seem to be a more relevant quantity for experimental observation. Of course, the relative magnitude of the global fluctuations is inversely proportional to the sample area $L_xL_y$, so the sample should be small enough.
We assume $L_x,L_y\gg{d}$ so that one can neglect boundary effects at the sample edges. At the same time, we assume $L_x,L_y\ll{c}/\omega_T$ so that the Coulomb approximation is valid. In practice, for $\omega_T\sim10^{14}\:\mbox{s}^{-1}$ we have $c/\omega_T\sim10\:\mu\mbox{m}$, while the separation can be $d\sim10\:\mbox{nm}$, so these assumptions are quite realistic.

\begin{figure}
    \centering
    \includegraphics[width=0.95\linewidth]{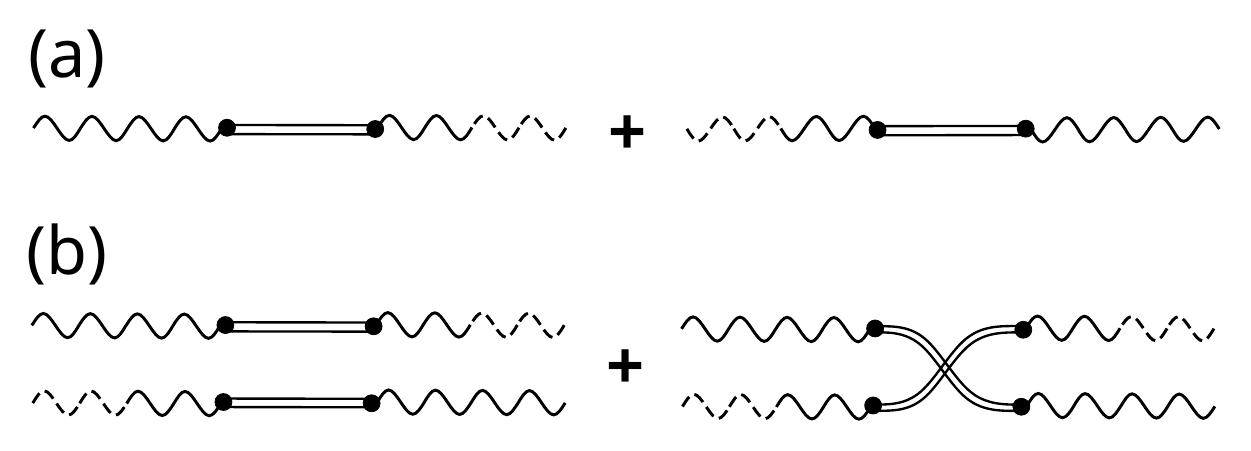}
    \caption{Feynamnn diagrams representing (a)~the average Poynting vector, and (b)~its fluctuations. Double solid lines represent the polarization operator $\tilde\Pi$, solid wavy lines represent the electric field Green's function~$D$, while dashed-solid and solid-dashed wavy lines represent the magnetic field response $-i(\omega/c)\Lambda$ and $i(\omega/c)\Lambda^\dagger$, respectively.}
    \label{fig:diag_noise}
\end{figure}

Technically, we need to calculate various averages of electric and magnetic field operators in the Heisenberg representation. It is convenient to rewrite Eqs.~(\ref{eqs:E=b=}) as
\begin{subequations}\begin{align}
    \hat{E}_i(\vec{r}) {}&{} = -\int{d}^3\vec{r}'\,D_{ik}(\vec{r}-\vec{r}')\,\hat{P}_k(\vec{r}') \nonumber\\ {}&{} = -\int{d}^3\vec{r}'\,\hat{P}_k(\vec{r}')\,D_{ki}(\vec{r}'-\vec{r}),\\
    \hat{B}_i(\vec{r}) {}&{} = -\frac1c\int{d}^3\vec{r}'\,\Lambda_{ik}(\vec{r}-\vec{r}')\,\hat{j}_k(\vec{r}') \nonumber\\ {}&{} = -\frac1c\int{d}^3\vec{r}'\,\hat{j}_k(\vec{r}')\,\Lambda_{ki}(\vec{r}'-\vec{r}).
\end{align}\end{subequations}
These linear relations between the fields and the polarization $\hat{\vec{P}}$ (recalling that the current $\hat{\vec{j}}=\partial\hat{\vec{P}}/\partial{t}$) enable us to express correlators of the fields in terms of correlators of the polarization given by Eqs.~(\ref{eqs:tildePi=}), as shown diagrammatically in Fig.~\ref{fig:diag_noise}.

To simplify the algebra keeping track of different combinations of the fields (giving 4 terms for $\langle\hat{J}_i(\vec{r},t)\rangle$ and 32 terms for $\langle\{\hat{J}_i(\vec{r},t),\hat{J}_k(\vec{r}',t')\}\rangle$), we arrange all components of the fields in a single 6-vector $(f_1,\ldots,f_6) \equiv (E_x,E_y,E_z,B_x,B_y,B_z)$ and consider a generic observable
\begin{equation}
    \hat{O}(\vec{r})=\sum_{\alpha,\beta=1}^6O_{\alpha\beta}\,\hat{f}_\alpha(\vec{r})\,\hat{f}_\beta(\vec{r}),
\end{equation}
with some matrix $O_{\alpha\beta}=O_{\beta\alpha}^*$. We also combine the two $3\times3$ matrices $D_{jk}(\vec{r}-\vec{r}')$ and $-i(\omega/c)\Lambda_{jk}(\vec{r}-\vec{r}')$, which determine response of the electric and magnetic fields to the external polarization, into a single $6\times3$ matrix $D_{\alpha{j}}(\vec{r},\vec{r},\omega)$. Then the average and the fluctuations of $\hat{O}(\vec{r},t)$, shown diagrammatically in Fig.~\ref{fig:diag_noise}, can be written as
\begin{widetext}\begin{subequations}\label{eqs:Oav}\begin{align}
&\langle\hat{O}(\vec{r},t)\rangle = \frac{i\hbar}2\sum_{\alpha,\beta}O_{\alpha\beta}\int{d}^3\vec{r}_1\,d^3\vec{r}_2\,\int\frac{d\omega}{2\pi}\,
D_{\beta{j}}(\vec{r},\vec{r}_1,\omega)\,
\tilde\Pi^{K-R+A}_{jk}(\vec{r}_1,\vec{r}_2,\omega)\,
{D}^*_{k\alpha}(\vec{r}_2,\vec{r}',\omega),\\
&\int{d}(t-t')\,e^{i\Omega(t-t')}\left[\frac12\langle\{\hat{O}(\vec{r},t),\hat{O}(\vec{r}',t')\}\rangle
- \langle\hat{O}(\vec{r},t)\rangle\langle\hat{O}(\vec{r}',t')\rangle\right]
\nonumber\\ &{} 
= -\frac{\hbar^2}4\sum_{\alpha,\beta,\alpha',\beta'}
\left(O_{\alpha\beta}O_{\alpha'\beta'}+O_{\alpha\beta}O_{\beta'\alpha'}\right)
\int{d}^3\vec{r}_1\ldots{d}^3\vec{r}_4\,\int\frac{d\omega}{2\pi}
\nonumber\\ &{} \quad \times
\left[{D}_{\alpha{i}}(\vec{r},\vec{r}_1,\omega+\Omega)\,
\tilde\Pi^K_{ij}(\vec{r}_1,\vec{r}_2,\omega+\Omega)\,
{D}^*_{j\alpha'}(\vec{r}_2,\vec{r}',\omega+\Omega)\,
{D}_{\beta'{k}}(\vec{r}',\vec{r}_3,\omega)\,
\tilde\Pi^K_{kl}(\vec{r}_3,\vec{r}_4,\omega)\,
{D}^*_{l\beta}(\vec{r}_4,\vec{r},\omega)
\right.\nonumber\\ {}&{}\qquad{}-{}\left.
\left(\tilde\Pi^K\to\tilde\Pi^R-\tilde\Pi^A\right)\right],
\end{align}\end{subequations}\end{widetext}
where we introduced a compact notation $\tilde\Pi^{K-R+A}=\tilde\Pi^{K}-(\tilde\Pi^{R}-\tilde\Pi^{A})$. Instead of the dressed polarization operators $\tilde\Pi$ and the empty-space field Green's functions~$D$, it is convenient to work with the mechanical polarization operators~$\Pi$ and the dressed field Green's functions~$\mathcal{D}$. Similarly to  Eqs.~(\ref{eq:tildePiRA=}),~(\ref{eq:tildePiK=}), in Eqs.~(\ref{eqs:Oav}) we can replace
\begin{subequations}\label{eqs:DPiD}\begin{align}
     &D_{\alpha{i}}\tilde\Pi^K_{ij}D_{j\beta}^* = \mathcal{D}^R\Pi^K_{ij}\mathcal{D}^A_{j\beta},\\
     &D_{\alpha{i}}(\tilde\Pi^R_{ij}-\tilde\Pi^A_{ij})D_{j\beta}^* = \mathcal{D}^R(\Pi^R_{ij}-\Pi^A_{ij})\mathcal{D}^A_{j\beta}.
\end{align}\end{subequations}
The mechanical polarization operators have a very simple structure~(\ref{eqs:PiKRA=}), while the dressed field Green's function~$\mathcal{D}^R_{\alpha{j}}$ can be found from Maxwell's equations in the given dielectric structure.

In the planar geometry we are considering here (vacuum for $|z|<d/2$, spatially uniform dielectric for $|z|>d/2$), it is natural to work in the Fourier space with respect to the in-plane coordinates $\vec{r}_\|\equiv(x,y)$, introducing the in-plane wave vector~$\vec{q}$. Let us place a layer of an external polarization at $z=z'<-d/2$, so that
\begin{equation}
\vec{P}^\text{ext}(\vec{r},t) = \vec{P}e^{i\vec{q}\vec{r}_\|-i\omega{t}}\delta(z-z').    
\end{equation}
Then, Maxwell's equations in the quasi-static limit $c\to\infty$ give the electrostatic potential~$\varphi$ together with the electric and magnetic fields in the region $|z|<d/2$:
\begin{subequations}\begin{align}
\varphi(\vec{r},t) ={}&{}  \frac{4\pi(i\vec{q}\vec{P}/q-P_z)e^{q(z'+d/2)+i\vec{q}\vec{r}_\|-i\omega{t}}}{(\varepsilon+1)^2e^{qd}-(\varepsilon-1)^2e^{-qd}}
\nonumber\\ {}&{} \times
\left[(\varepsilon-1)e^{q(z-d/2)} - (\varepsilon+1)e^{-q(z-d/2)}\right],\\
\vec{E}(\vec{r},t) = {}&{} i\vec{q}\,\frac{4\pi(i\vec{q}\vec{P}/q-P_z)e^{q(z'+d/2)+i\vec{q}\vec{r}_\|-i\omega{t}}}{(\varepsilon+1)^2e^{qd}-(\varepsilon-1)^2e^{-qd}}
\nonumber\\ {}&{} \quad\times
\left[(\varepsilon+1)e^{-q(z-d/2)}-(\varepsilon-1)e^{q(z-d/2)}\right] \nonumber\\
{}&{} + q\vec{e}_z\,\frac{4\pi(i\vec{q}\vec{P}/q-P_z)e^{q(z'+d/2)+i\vec{q}\vec{r}_\|-i\omega{t}}}{(\varepsilon+1)^2e^{qd}-(\varepsilon-1)^2e^{-qd}}
\nonumber\\ {}&{} \quad\times
\left[(\varepsilon+1)e^{-q(z-d/2)}+(\varepsilon-1)e^{q(z-d/2)}\right],\\
\vec{B}(\vec{r},t) = {}&{} \frac\omega{c}\,\frac{\vec{e}_z\times\vec{q}}q\,\frac{4\pi(i\vec{q}\vec{P}/q-P_z)e^{q(z'+d/2)+i\vec{q}\vec{r}_\|-i\omega{t}}}{(\varepsilon+1)^2e^{qd}-(\varepsilon-1)^2e^{-qd}}
\nonumber\\ {}&{} \times
\left[(\varepsilon+1)e^{-q(z-d/2)}+(\varepsilon-1)e^{q(z-d/2)}\right],
\end{align}\end{subequations}
where $\vec{e}_z$ is the unit vector along the $z$~axis. Thus, if for each given~$\vec{q}$ we rotate the basis $(x,y,z)\to(L,T,Z)$ with unit vectors $\vec{e}_L=\vec{q}/q$, $\vec{e}_T=\vec{e}_z\times\vec{q}/q$, $\vec{e}_Z=\vec{e}_z$, the field Green's function $\mathcal{D}^R_{\alpha{j}}$ has a direct product structure with only 4 non-zero components out of~18:
\begin{subequations}\begin{align}
&\mathcal{D}^R(-d/2<z<d/2,z'<-d/2,\vec{q},\omega)  = {}\nonumber\\
{}&{} 
= \frac{4\pi{e}^{q(d/2+z')}}{\nu_+(0)\,\nu_-(0)}
\begin{pmatrix} iq\nu_-(z) \\ 0 \\ q\nu_-(z) \\ 0 \\ (\omega/c)\nu_+(z) \\ 0\end{pmatrix}
\begin{pmatrix} -i & 0 & 1\end{pmatrix} ,\\
&\nu_\pm(z)\equiv(\varepsilon+1)e^{-q(z-d/2)}\pm(\varepsilon-1)e^{q(z-d/2)}.
\end{align}
If the source polarization is placed at $z'>d/2$, the fields are easily found from the previous solution by symmetry: flipping the signs $z\to-z$, $z'\to-z'$ implies  $(P_x,P_y,P_z)\to(P_x,P_y,-P_z)$, $(E_x,E_y,E_z)\to(E_x,E_y,-E_z)$, $(B_x,B_y,B_z)\to(-B_x,-B_y,B_z)$:
\begin{align}
&\mathcal{D}^R(-d/2<z<d/2,z'>d/2,\vec{q},\omega) = {}\nonumber\\
{}&{} = \frac{4\pi{e}^{q(d/2-z')}}{\nu_+(0)\,\nu_-(0)}
\begin{pmatrix} -iq\nu_-(-z) \\ 0 \\ q\nu_-(-z) \\ 0 \\ (\omega/c)\nu_+(-z) \\ 0\end{pmatrix}
\begin{pmatrix} i & 0 & 1\end{pmatrix}.
\end{align}\end{subequations}
For $z=d/2-0^+$ we are mainly interested in, we have simply $\nu_+(d/2)=2\varepsilon$, $\nu_-(d/2)=2$. 

For a finite-size sample, the in-plane momentum $\vec{q}$ takes discrete values, spaced by $\pi/L_x,\pi/L_y$. However, for $L_x,L_y\gg{d}$ we can neglect this discretization and replace discrete summation $\sum_{\vec{q}}\to{L}_xL_y\int{d}^2\vec{q}/(2\pi)^2$. The matrix $O_{\alpha\beta}$ corresponding to the Poynting vector has the same simple form in the original $(x,y,z)$ basis and in the rotated $(L,T,Z)$ basis (since $\hat{J}_z$ is invariant under rotations around the $z$~axis):
\begin{equation}
O = \frac{c}{8\pi}\begin{pmatrix}
0 & 0 & 0 & 0 & 1 & 0  \\ 0 & 0 & 0 & -1 & 0 & 0 \\
0 & 0 & 0 & 0 & 0 & 0 \\ 0 & -1 & 0 & 0 & 0 & 0 \\
1 & 0 & 0 & 0 & 0 & 0 \\ 0 & 0 & 0 & 0 & 0 & 0
\end{pmatrix}.
\end{equation}
All matrix products, spatial convolutions and $z$ integrations in Eqs.~(\ref{eqs:Oav}) with replacements~(\ref{eqs:DPiD}) can be straightforwardly performed now.

For the average power, the result is
\begin{align}
\frac{\langle\hat{\cal P}_1\rangle}{L_xL_y} ={}&{}  \int\frac{d^2\vec{q}}{(2\pi)^2}\int_0^\infty\frac{d\omega}{2\pi}\,\left(\frac{\hbar\omega}{e^{\hbar\omega/T_2}-1}-\frac{\hbar\omega}{e^{\hbar\omega/T_1}-1}\right)
\nonumber\\
{}&{}\times\frac{16(\Im\varepsilon)^2}{|(\varepsilon+1)^2e^{qd}-(\varepsilon-1)^2e^{-qd}|^2}.
\end{align}
Using Eq.~(\ref{eq:wpmq=}), we can rewrite the last line identically in terms of the surface mode frequencies $\omega_\pm(q)$ as
\begin{align}
& |S_{12}(\vec{q},\omega)|^2\equiv \frac{16(\Im\varepsilon)^2}{|(\varepsilon+1)^2e^{qd}-(\varepsilon-1)^2e^{-qd}|^2} \nonumber\\ 
&{} = \frac{(2\gamma\omega)^2[\omega_+^2(q)-\omega^2_-(q)]^2}{[(\omega_+^2(q)-\omega^2)^2+(2\gamma\omega)^2][(\omega_-^2(q)-\omega^2)^2+(2\gamma\omega)^2]},
\end{align}
which is identical to Eq.~(\ref{eq:S21modsquare}) for the effective circuit. In fact, $|S_{12}(\vec{q},\omega)|^2$ appearing here is identical to $|S_{12}(\omega)|^2$ in Eq.~(\ref{eq:S21modsquare}) for any frequency dependence of $\varepsilon(\omega)$, if one makes a substitution
\begin{equation}\label{eq:substitution}
    Z(\omega)\to-\frac{1}{i\omega{C}_*}\,\frac{1}{\varepsilon(\omega)},\quad
    C_\pm\to{C}_*\,\frac{e^{qd}\mp1}{e^{qd}\pm1},
\end{equation}
where $C_*$ is an arbitrary constant having the dimensionality of capacitance.

For the correlator of the power $\hat{\mathcal{P}}_1$, we find [denoting $\omega'\equiv\omega+\Omega$, and the prime at $\nu_\pm'(z)$ indicating that it contains $\varepsilon(\omega')$]
\begin{widetext}
\begin{align}
 \frac{W_{11}(\Omega)}{L_xL_y} 
{}&{}= \int_{-\infty}^\infty\frac{d\omega}{16\pi}\int\frac{d^2\vec{q}}{(2\pi)^2} \, \frac{2\hbar\Im\varepsilon(\omega')}{|\nu_+'(0)\,\nu_-'(0)|^2}\, \frac{2\hbar\Im\varepsilon(\omega)}{|\nu_+(0)\,\nu_-(0)|^2} \times {}\nonumber\\
{}&{}\times\left[\left(\coth\frac{\hbar\omega'}{2T_2}\coth\frac{\hbar\omega}{2T_2}-1\right)\left|\nu_-^*(d/2)\,\omega'\nu_+'(d/2)-\omega\nu_+^*(d/2)\,\nu_-'(d/2)\right|^2
\right. + \nonumber\\ {}&{} \quad + \left.
\left(\coth\frac{\hbar\omega'}{2T_1}\coth\frac{\hbar\omega}{2T_2}-1\right)
\left|\nu_-^*(d/2)\,\omega'\nu_+'(-d/2)+\omega\nu_+^*(d/2)\,\nu_-'(-d/2)\right|^2
\right. + \nonumber\\ {}&{} \quad + \left.
\left(\coth\frac{\hbar\omega'}{2T_2}\coth\frac{\hbar\omega}{2T_1}-1\right)
\left|\nu_-^*(-d/2)\,\omega'\nu_+'(d/2)+\omega\nu_+^*(-d/2)\,\nu_-'(d/2)\right|^2\right. + \nonumber\\ {}&{} \quad + \left.
\left(\coth\frac{\hbar\omega'}{2T_1}\coth\frac{\hbar\omega}{2T_1}-1\right)
\left|\nu_-^*(-d/2)\,\omega'\nu_+'(-d/2)-\omega\nu_+^*(-d/2)\,\nu_-'(-d/2)\right|^2\right].
\end{align}
\end{widetext}
Again, at each $\vec{q}$ the integrand of this expression is identical to Eq.~\eqref{eq:Wjk=} if one takes the scattering matrix~(\ref{eqs:SmatrixPolariton=}) and makes the substitution~(\ref{eq:substitution}).
That is, each in-plane momentum~$\vec{q}$, which takes discrete values for finite $L_x,L_y$, represents an independent scattering channel admitting an equivalent circuit representation. Since for $q\gg1/d$ the integrand is exponentially suppressed, the effective number of channels contribution to the heat current and its fluctuations is of the order of $L_xL_y/d^2$. Thus, a single circuit can qualitatively represent heat transfer between two pieces of dielectric whose size $L_x,L_y\sim{d}$, where the number of channels is of the order of~1.

\bibliography{Bib}

\end{document}